\documentclass[aps,prd,nofootinbib,twocolumn,groupedaddress, showpacs]{revtex4}

\usepackage{amssymb,amsmath}
\usepackage{epsf}
\usepackage{graphicx}
\usepackage{xcolor}
\usepackage{ulem}

\def\mprp{\mbox{\tiny $\bot$}}
\def\mprl{\mbox{\tiny $\|$}}

\def\beq{\begin{eqnarray}}
\def\eeq{\end{eqnarray}}
\def\ee{\varepsilon}
\def\lm{\lambda}
\newcommand{\prp}[1]{#1_{\mbox{\tiny $\bot$}}}
\newcommand{\prll}[1]{#1_{\mbox{\tiny $\|$}}}
\def\HH{H\!\!\left(\frac{\prll{q}^2}{4 m^2}\right)}
\def\HHi{H\!\!\left(\frac{\prll{q'^2}}{4 m^2}\right)}
\def\HHii{H\!\!\left(\frac{\prll{q''^2}}{4 m^2}\right)}

\def\1{1 \to 1 \, 2}
\def\2{1 \to 2 \, 2}

\def\A{{\cal A}}
\def\P{{\cal P}}
\def\M{{\cal M}}

\begin{document}

\title{Photon splitting and Compton scattering in strongly magnetized hot plasma}

\author{M. V. Chistyakov} \email[]{mch@uniyar.ac.ru}
\author{D. A. Rumyantsev} \email[]{rda@uniyar.ac.ru}
\author{N. S. Stus'}
\affiliation{Division of Theoretical Physics, Yaroslavl State (P.G.~Demidov) University, Sovietskaya 14, 150000 Yaroslavl, Russia (Russian Federation)}

% \author{M. V. Chistyakov, D. A. Rumyantsev \\[3mm]
% {\small\it Division of Theoretical Physics,} \\
% {\small\it Yaroslavl State (P.G.~Demidov) University,} \\
% {\small\it Sovietskaya 14, 150000 Yaroslavl, Russian Federation}\\
% {\small\tt E-mail: mch@uniyar.ac.ru, rda@uniyar.ac.ru}
% }
\date{\today}

\begin{abstract}
The process of photon splitting is investigated in the presence
of strongly magnetized electron-positron plasma.
The amplitude of the process is calculated in the general case of plasma with nonzero 
chemical potential and temperature. 
%We calculate the amplitude of the process at the
%nonzero chemical potential and temperature.
The polarization selection rules and corresponding partial amplitudes for 
allowed splitting channels are obtained 
in the case of charge-symmetric plasma.
It is found that the new splitting channel forbidden in magnetized vacuum
becomes allowed.
The absorption rates of the
photon splitting are calculated with taking into account  the photon
dispersion and wave function renormalization. 
In addition, the comparison of photon splitting and Compton scattering 
process is made. The influence of the reactions under consideration on the 
radiation transfer in the framework
of the magnet ar model of  a
soft gamma repeater burst is discussed.
\end{abstract}


\maketitle

\section{Introduction}

The process of photon splitting into two photons is a prominent
example of the external active medium influence on the reactions
with  elementary particles. Though it is forbidden in vacuum by
charged conjugation symmetry of quantum electrodynamics (QED), known as Furry's theorem, it
becomes allowed in the presence of an external electromagnetic field
and/or plasma. In spite of the rather long history of the
investigation
~\cite{Adler:1970,Bialynicka:1970,Adler:1971,Ritus:1972,Stoneham:1979,
Mentzel:1994,Adler:1996,Baier:1996,Chistyakov:1998,Weise:2004} (for
 a review see e.g.~\cite{Papanian:1989,LaiHarding:2006}) the process of
photon splitting still attracts great interest concerning its
possible applications.

It is remarkable, that such an exotic process may lead to observable physical
manifestations in astrophysical objects.
Its principal effect is to degrade photon
energies and thereby soften gamma-ray spectra from neutron stars.
In particular it is supposed that photon splitting could explain a
peculiarity in the gamma-ray spectrum of some
radio pulsars~\cite{HBG:1997}.

Soft gamma repeaters (SGRs) and anomalous x-ray
pulsars (AXPs) belong to the another class of  astrophysical objects
where the process of photon splitting could play an important role. There
is a number of arguments that these objects are magnetars, a
distinct class of neutron stars with magnetic field strength  of
$B \sim 10^{14}-10^{16}$ G~\cite{Duncan:1992}, i.e. $B \gg B_e$, where $B_e = m^2/e \simeq 4.41
\times 10^{13}$~G~\footnote{We use natural units $c = \hbar = k = 1$, $m$
is the electron mass, and   $e > 0$ is the elementary charge.} is the critical magnetic field.
It has been suggested that
the degrading action of $\gamma\to\gamma\gamma$ could be responsible
for the spectral cutoffs in the spectra of SGRs~\cite{Baring:1995, Thompson:1995}.

There is one more interesting application of the process under consideration
concerning the radio quiescence of SGRs and AXPs.
Because photon splitting has no threshold,
high-energy photons propagating at very small angles to the magnetic
field in neutron star magnetospheres may split before reaching the
threshold of pair production.  Therefore, this process could change
the production efficiency of electron-positron pairs required for a detectable
radio emission~\cite{BH:1998, Malofeev:2005, Istomin:2007}.

The process discussed here also plays an important role in the model of SGR burst~\cite{Thompson:1995, Thompson:2001}. It is assumed 
the formation of the magnetically "trapped fireball" in the near magnetosphere of a magnetar with 
hot $e^{+} e^{-}$-photon plasma occurs in local thermodynamic equilibrium.   

Notice that in all of these astrophysical models the photon splitting process in
the presence of strong magnetic field is considered in the environment which is non empty space. Instead the presence of
considerably dense and hot electron-positron plasma is possible. This plasma
could significantly change the photon splitting rate.

There are two ways in which a magnetized plasma influences the process
under consideration.  On the one hand, it modifies the photon
dispersion properties. On the other hand it can change the process amplitude.
The  first effect has been studied previously in 
Refs.~\cite{Adler:1971, Bulik:1998}.
It was shown that  in the case of the cold
weakly magnetized plasma the process kinematics and polarization
selection rules remain the same as in the magnetized vacuum
if  the electron density is not too large
($n_e \lesssim 10^{19} cm^{-3}$)~\cite{Adler:1971}. In~\cite{Bulik:1998}
the probability  of the photon splitting  was calculated by
including the plasma effects on the photon dispersion but using the
process amplitude obtained in the weak magnetic field without plasma.
It was found that the plasma influence in such an approach was
negligibly small except the very narrow region of plasma and magnetic field
parameter space.

The modification of the photon splitting amplitude in the presence of the magnetic
field and plasma  was considered
in~\cite{Elmfors:1998, Gies:2000}  on the basis of the
Euler-Heisenberg effective Lagrangian with taking account of
thermal corrections in the one-loop and two-loop approximations correspondingly.
It was noted that in the low-temperature limit the process
$\gamma \to \gamma \gamma$ could compete with the other  absorption reactions
such as the Compton scattering process.

Another approach was applied in~\cite{Martinez:2001}. Using the
expansion of the electron propagator over of the magnetic field
strength the amplitude and the absorption coefficient of photon
splitting were calculated in the high-energy limit. The main conclusion
of the paper was that the plasma effect was negligible. However, the
estimations of the photon splitting absorption coefficients  in
the magnetic field presented in the paper in the high energy limit were
incorrect because these expressions were applicable only in the low-energy approximations. Note that in~\cite{Elmfors:1998, Gies:2000,
Martinez:2001} the effects of the photon dispersion in magnetized
plasma were not taken into account.

Unlike in the pure magnetic field, in the presence of plasma the photon
can additionally  scatter directly on electrons and positrons
(Compton scattering). Moreover, in hot plasma the inverse process of
photon merging, $\gamma \gamma \to \gamma$ also occurs. To be
consistent, one has to compare all  these processes under the same
conditions. Previously, the Compton scattering in magnetic field
without plasma was investigated in~\cite{Herold:1979, Melrose:1983,
Harding:1986, Harding:2000}.  It was found that it  became strongly
anisotropic and essentially depended on a photon polarization.
In~\cite{Bulik:1997}, it was shown that the dispersion properties of
a photon in  strongly magnetized cold plasma could significantly
influence  the Compton scattering. The process of photon merging
was considered in~\cite{Thompson:1995} as one of the dominant
number-photon changing process. The case of low energies and a strong
magnetic field without plasma was investigated there.

Therefore,  one can conclude that no self-consistent investigation
of  the photon splitting/merging and Compton scattering  including
modification of both  dispersion relations and the process amplitude
in magnetized plasma is currently available. Moreover, the case of
the strong magnetic field, $B \gtrsim B_e$, which is relevant for
the astrophysical applications has not been investigated for the
process of photon splitting.

In the present paper we would like to fill in this gap. The processes
of photon splitting/merging, $\gamma \leftrightarrow \gamma \gamma$
and  Compton scattering are considered in the case of  strongly
magnetized $e^+e^-$-plasma, when the magnetic field strength $B$ is the
maximal physical parameter, namely $\sqrt{eB} \gg T,\, \mu,\,
\omega$. Here, $T$ is the plasma temperature, $\mu$ is the chemical
potential, $\omega$ is the initial photon energy. Under these
conditions, electrons and positrons in plasma occupy mainly the
ground Landau level. The more accurate relation for the magnetic field
and plasma parameters in this case can be obtained from the condition that the energy 
density of magnetic field must be much larger than plasma energy density~\cite{Mikheev:2000}. 
For example, in ultrarelativistic plasma this condition leads to the following relation:\ 
$$
\frac{B^2}{8 \pi} \gg \frac{\pi^2 (n_{e^-} - n_{e^+})^2}{eB} +
\frac{eB T^2}{12},
$$
where $n_{e^-}, n_{e^+}$ are the electron and positron number densities.

The paper is organized as follows. In Sec.~\ref{Sec:2} we
 calculate   the photon splitting amplitude in the
strong magnetic field and plasma in the general case of nonzero chemical potential and 
temperature. Taking in mind the possible application of our results to the magnetar model of SGR burst in 
all the following sections the case of the charge-symmetric 
electron-positron plasma ($\mu = 0 $) is considered. In  Sec.~\ref{Sec:3} we
perform an analysis of the $\gamma \to \gamma \gamma$  kinematics and
consider the photon polarization selection rules. A numerical
calculation of the photon splitting and merging absorption rates for
different channels are presented in Sec.~\ref{Sec:4}. In
Sec.~\ref{Sec:5} we consider the process of Compton scattering
and compare it with the photon splitting. Final comments and discussion
of the obtained results and possible astrophysical applications are given in
Sec.~\ref{Sec:6}.
\section{Photon splitting amplitude}\label{Sec:2}
\begin{figure*}
\includegraphics[width=13.5cm]{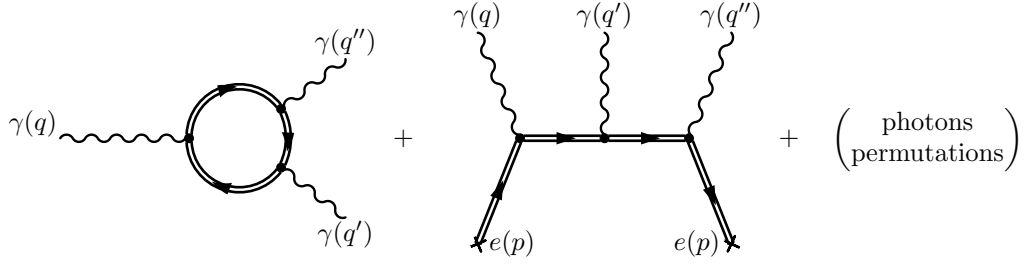}
\caption{ The Feynman diagrams for the photon splitting process in
magnetized plasma. The cross at the end of the electron line
symbolizes that the particle belongs to the medium. }
\label{fig:DiagGGG}
\end{figure*}
The amplitude of the process under consideration  can be presented
as the sum of two
terms,
\begin{eqnarray}
  {\cal M}\, = \, {\cal M}_B + {\cal M}_{pl},
\label{eq:M1}
\end{eqnarray}

\noindent where ${\cal M}_B$ is the amplitude in the magnetized
vacuum~\footnote{The details of the  calculation of 
the amplitude  can be found in Ref.~\cite{RCh05}.}. The second term in (\ref{eq:M1}) is the plasma contribution
which can  be treated as the photon coherent scattering from the real
electrons and positrons with two photons emissions without change of
their states (''forward`` scattering). These contributions are
depicted by the Feynman diagrams in  Fig.~\ref{fig:DiagGGG},
\noindent where the internal double lines correspond to the electron
propagator in a magnetic  field. In the case of strongly
magnetized plasma it is relevant to use the propagator in  the asymptotic
form~\cite{Chistyakov:2002}:
\begin{eqnarray}
S(x,y) &=& e^{i \Phi(x,y)} \hat S(x-y),
\label{eq:S}
\end{eqnarray}
where
\begin{eqnarray}
\hat S(X) &=& \hat S_{-}(X) + \hat S_{+}(X) +  \prp{\hat S}(X),
\\
\hat S_{-}(X) &\simeq&
\frac{i eB}{2 \pi} \exp \left (- \frac{eB \prp{X}^2}{4}\right )
\nonumber \\
&\times&
\int \frac{d^2 p}{(2 \pi)^2}\, \frac{(p\gamma)_{\mbox{\tiny $\|$}} + m}{\prll{p}^2 - m^2}
\Pi_{-}e^{-i \prll{(pX)}},
\label{eq:S-a}
\\
\hat S_{+}(X)  &\simeq&
-\frac{i}{4 \pi}\,
[i\prll{(\gamma \,\partial/\partial X)} + m]\,
\prll{\delta}^2(X) \, \Pi_{+}
\nonumber
\\
&\times&
\exp \left (\frac{eB \prp{X}^2}{4}\right )\,
\Gamma \left (0, \frac{eB \prp{X}^2}{2} \right ),
\label{eq:S+a}
\end{eqnarray}
\begin{eqnarray}
\prp{\hat S}(X) &\simeq&
- \frac{1}{2 \pi} \, \prll{\delta}^2(X)\, \frac{\prp{(X \gamma)}}{\prp{X}^2}
\exp \left (- \frac{eB \prp{X}^2}{4}\right ).
\label{eq:S_pr_a}
\\
d^2 p &=& dp_0 dp_3, \quad \Pi_{\pm} = \frac{1}{2} (1 \pm i \gamma_1
\gamma_2),
\nonumber \\
\Pi^2_\pm &=& \Pi_\pm, \quad \quad  \quad [\Pi_\pm, \prll{(A \gamma)}] = 0,
\nonumber
\end{eqnarray}
Here, $\gamma_\alpha$ are the Dirac matrices in the standard representation,
$\Gamma(a, z)$ is the incomplete gamma function,
$
\Gamma(a,z) = \int \limits^\infty_z t^{a - 1} e^{-t} dt,
$ and
\begin{eqnarray}
\Phi(x,y) = - e \int \limits_x^y d \xi_\mu \left[ \A_\mu(\xi) +
\frac{1}{2} F_{\mu \nu} (\xi - y)_\mu \right ],
\label{eq:Phi}
\end{eqnarray}
where $\A_\mu$ and $F_{\mu \nu}$ are the 4-potential and  the tensor of the
uniform magnetic field, correspondingly. Hereafter we use the notation:
4-vectors with the indices $\bot$ and $\parallel$
belong
to the Euclidean (1, 2) subspace and the Minkowski (0, 3) subspace correspondingly, when the
field $\bf B$ is directed along the third axis. Then for arbitrary 4-vectors
$A_\mu$, $B_\mu$ one has
\begin{eqnarray}
&&\prp{A}^\mu = (0, A_1, A_2, 0), \quad  \prll{A}^\mu = (A_0, 0, 0, A_3), \nonumber \\
&&\prp{(A B)} = (A \Lambda B) =  A_1 B_1 + A_2 B_2 , \nonumber \\
&&\prll{(A B)} = (A \widetilde \Lambda B) = A_0 B_0 - A_3 B_3, \nonumber
\end{eqnarray}

\noindent where the matrices
$\Lambda_{\mu \nu} = (\varphi \varphi)_{\mu \nu}$,\,
$\widetilde \Lambda_{\mu \nu} =
(\tilde \varphi \tilde \varphi)_{\mu \nu}$ are constructed with
the dimensionless tensor of the external
magnetic field, $\varphi_{\mu \nu} =  F_{\mu \nu} /B$,
and the dual one,
${\tilde \varphi}_{\mu \nu} = \frac{1}{2}
\varepsilon_{\mu \nu \rho \sigma} \varphi_{\rho \sigma}$.
Matrices $\Lambda_{\mu \nu}$ and  $\widetilde \Lambda_{\mu \nu}$
are connected by the relation
$\widetilde \Lambda_{\mu \nu} - \Lambda_{\mu \nu} =
g_{\mu \nu} = diag(1, -1, -1, -1)$,
and play the role of the metric tensors in the perpendicular ($\bot$)
and  parallel ($\parallel$) subspaces respectively.

The double external lines in Fig. 1 correspond to the
solution of the Dirac equation in a magnetic field for
the electrons  on the ground Landau level. For the choice of
gauge $\A_\mu=(0, 0, B x, 0),$ they are
\begin{eqnarray}
\psi_{p\, \epsilon} = \frac{(eB)^{1/4}}{(\sqrt{\pi} 2 E L_y
L_z)^{1/2}} e^{-i \epsilon (E t - p_y y - p_z z)}
e^{-\xi_\epsilon^2/2} u_\epsilon(\prll{p}), \label{eq:Dirac_solution}
\end{eqnarray}
where $\xi_\epsilon = \sqrt{eB} (x + \epsilon \frac{p_y}{eB})$,
$\epsilon = \pm 1$ denotes the solutions for electron with positive
and negative energy correspondingly,  $E = \sqrt{p_z^2 + m^2}$. The
bispinor amplitudes are given by
\begin{eqnarray}
u_\epsilon(\prll{p}) = \frac{1}{\sqrt{E + \epsilon \, m}} \left (
\begin{array}{c}
(E + \epsilon \,m) \Psi \\
- p_z \Psi
\end{array}
\right ), \,
\Psi = \left (
\begin{array}{c}
0\\
1
\end{array}
\right ).
% \nonumber
% \\
\end{eqnarray}
Then the plasma contribution can be obtained by the summation
of the scattering diagrams in Fig. 1 over all electron and positron states
with taking into account  their distribution function in plasma.
After rather lengthy but straightforward calculation we
obtain the following expression for the amplitude:
\beq
\M = \, \ee_\mu(q) \ee^*_\nu(q'') \ee^*_\rho(q') \,\Pi_{\mu \nu \rho},
\label{eq:M}
\eeq
\noindent where
\begin{widetext}
\beq
\Pi_{\mu \nu \rho} &=& eB \,\frac{e^3}{4 \pi^2}
\frac{(\tilde \varphi q)_\mu
(\tilde \varphi q'')_\nu (\tilde \varphi q')_\rho}
{(q' \tilde \varphi q'')}
\Big [{\cal J}_{2}^{(-)}(\prll{q},\prll{q'}) \, - \,
{\cal J}_{2}^{(-)}(\prll{-q'},\prll{-q}) \, - \,
% \nonumber \\[3mm]
% &-&
{\cal J}_{2}^{(-)}(\prll{-q''},\prll{q'}) \, - \, (q' \leftrightarrow q'')
\Big]  
\label{eq:Pi0}\\
% \eeq
% \end{widetext}
%
% \begin{widetext}
% \beq
 &-& \frac{ i e^3}{2 \pi^2} \{ (q' \varphi q'') \,
[\pi_{\mu\nu \rho} + \upsilon_{\mu\nu \rho}] +
(q' {\cal G}(q''))_\nu \,\varphi_{ \rho \mu} +
% \nonumber \\[3mm]
% &+&
\frac{1}{2} ((q'' - q') {\cal G}(q))_\mu\,
\varphi_{ \nu \rho} +
(q'' {\cal G}(q'))_\rho \,\varphi_{ \nu \mu}
\nonumber \\[3mm]
&-&
 {\cal G}_{\nu \rho}(q'')\, (q' \varphi)_\mu+
 {\cal G}_{\mu \nu}(q'')\, (q \varphi)_\rho +
{\cal G}_{\mu \rho}(q')\, (q \varphi)_\nu -
{\cal G}_{\nu \rho}(q')\, (q'' \varphi)_\mu -
% \nonumber \\[3mm]
% &-&
{\cal G}_{\mu \nu}(q)\, (q'' \varphi)_\rho -
{\cal G}_{\mu \rho}(q)\, (q' \varphi)_\nu\
\nonumber \\[3mm]
&-& \frac{i(\tilde \varphi q)_\mu (\tilde \varphi q'')_\nu (\tilde
\varphi q')_\rho}{4 (q' \tilde \varphi q'')} [\prp{q'^2} +
\prp{q''^2} + (q'q'')_{\mprp}] [{\cal
J}_{2}^{(-)}(\prll{q},\prll{q'}) \, - \, {\cal
J}_{2}^{(-)}(\prll{-q'},\prll{-q}) -
% \nonumber \\[3mm]
% &-&
{\cal J}_{2}^{(-)}(\prll{-q''},\prll{q'}) \, - \, (q' \leftrightarrow q'')]
\}.
\nonumber
%\label{eq:Pi1}
\eeq
Here
\begin{eqnarray}
{\cal G}_{\mu\nu}(q) &=& \bigg (
\tilde \Lambda_{\mu \nu} - \frac{q_{\mbox{\tiny $\|$}\mu} \,
q_{\mbox{\tiny $\|$}\nu}}{\prll{q^{2}}}\bigg )\, {\cal G}(q_{\mbox{\tiny $\|$}}),
\\
 {\cal G}(q_{\mbox{\tiny $\|$}})&=& \left [ H\!\!\left(\frac{q_{\mbox{\tiny $\|$}}^2}{4 m^2}\right)  + {\cal J}_{1}(q_{\mbox{\tiny $\|$}}) \right ],
% \end{eqnarray}
% %
% \beq
\label{eq:Gq}
%\\[3mm]
 \end{eqnarray}
\begin{eqnarray}
{\cal J}_{1}(\prll{q}) &=& 2 \prll{q}^2 m^2 \int \frac{dp_z}{E} \,
\frac{f_{-}(E) + f_{+}(E)} {\prll{q}^4 - 4 \prll{(pq)}^2},
\label{eq:J1}
\\[3mm]
%\end{eqnarray}
%
%\begin{eqnarray}
{\cal J}_{2}^{(\pm)}(\prll{q},\prll{q'}) &=& 2 m^2 \int
\frac{dp_z}{E} \, \frac{f_{-}(E) \pm f_{+}(E)} {[q_{\mprl}^2 + 2
(pq)_{\mprl}][\prll{q'^2} + 2 (pq')_{\mprl}]},
\nonumber\\[2mm]
\label{eq:J2}
\end{eqnarray}
$f_{\mp}(E) \, = \, [e^{(E \, \mp \, \mu)/T} \, + \, 1]^{-1}$
 are the electron/positron distribution functions. The function $H(z)$ is defined as
\begin{eqnarray}
H(z) = \frac{1}{\sqrt{z(1 - z)}} \arctan{\sqrt{\frac{z}{1 - z}}} - 1,
\, z \le 1.
\end{eqnarray}

The expression for $\pi_{\mu\nu \rho}$ can be presented in the following
form
%
%\begin{widetext}
\begin{eqnarray}
\pi_{\mu \nu \rho} &=& \frac{1} {\prll{q^2} \prll{q'^2}
\prll{q''^2}} \Big [ (q' \tilde \varphi q'')\big\{ (\tilde \varphi
q)_\mu (\tilde \varphi q'')_\nu (\tilde \varphi q')_\rho \prp{\pi}
+(\tilde \varphi q)_\mu (\widetilde \Lambda q'')_\nu
( \widetilde \Lambda q')_\rho H
-(\widetilde \Lambda q)_\mu (\tilde \varphi q'')_\nu
( \widetilde \Lambda q')_\rho H''
 \nonumber
 \\
&-& (\widetilde \Lambda q)_\mu (\widetilde \Lambda q'')_\nu
( \tilde \varphi q')_\rho H'\big\}
+ \prll{(q' q'')}
( \widetilde \Lambda q)_\mu
(\tilde \varphi q'')_\nu (\tilde \varphi q')_\rho
(H'-H'')
+
\prll{(q q'')}
(\tilde \varphi q)_\mu (\tilde \varphi q'')_\nu
( \widetilde \Lambda q')_\rho
(H-H'')
\nonumber
\\
&+&
\prll{(q q')}
(\tilde \varphi q)_\mu
( \widetilde \Lambda q'')_\nu
(\tilde \varphi q')_\rho
(H'-H)
\Big ],
\end{eqnarray}
\begin{eqnarray}
\pi_{\mbox{\tiny $\bot$}} &=& H' + H'' + H
+ 2 \{ \prll{q^{2}} \prll{q'^2}\prll{q''^2}
- 2 m^2 [ \prll{q^2} \prll{(q'q'')} H - \prll{q'^2} \prll{(qq'')}
H' - \prll{q''^2} \prll{(qq')} H'']\} \nonumber
\\
&\times& \{\prll{q^{2}} \prll{q'^2}\prll{q''^2} - 4 m^2
[\prll{q'^2}\prll{q''^2} - \prll{(q'q'')^2}]\}^{-1}. \label{eq:pi}
\end{eqnarray}
Here,
$$
H \equiv \HH,\,
H' \equiv \HHi,\,
H'' \equiv \HHii.
$$
The expression for $\upsilon$ is
\begin{eqnarray}
\upsilon_{\mu\nu \rho}= \pi_{\mu\nu \rho}
\left [ \pi_{\mprp} \to \upsilon_{\mprp}, H \to {\cal J}_{1}(q_{\mprl}),
  H' \to {\cal J}_{1}(q'_{\mprl}), H'' \to {\cal J}_{1}(q''_{\mprl})\right],
\nonumber
\end{eqnarray}
\begin{eqnarray}
\prp{\upsilon} &=& \frac{1}{(q' \tilde \varphi q'')^2} \Big \{
(qq')_{\mprl}(qq'')_{\mprl}{\cal J}_{1}(q_{\mprl})
-(qq')_{\mprl}(q'q'')_{\mprl}{\cal J}_{1}(q'_{\mprl}) -
(qq'')_{\mprl}(q'q'')_{\mprl}{\cal J}_{1}(q''_{\mprl})
\nonumber
\\
&+& \frac{\prll{q^2} \prll{q'^2} \prll{q''^2}}{4} \Big [{\cal
J}_{2}^{(+)}(\prll{q},\prll{q'}) + {\cal
J}_{2}^{(+)}(\prll{-q'},\prll{-q}) 
+
{\cal J}_{2}^{(+)}(\prll{-q''},\prll{q'}) \, + \, (q' \leftrightarrow
q'')\Big] \Big \}.
\label{eq:upsilon}
\end{eqnarray}
\end{widetext}

As it was noted in~\cite{RCh05},  despite the fact that the 
amplitude (\ref{eq:M}) is obtained in the 
rest frame of plasma, it is possible  to generalize it
to the case when plasma moves as a whole along the magnetic
field. It is well-known that this is implemented by introducing the
velocity 4-vector $\upsilon_\mu$ ($\upsilon^2 = 1$) of the
medium~\cite{Fradkin:1965, Weldon:1982} in terms of which the
amplitude can be written in a covariant form.  In the presence of a 
magnetic field one should also introduce the condition for the
absence of an electric field which can be written as $\upsilon_\mu
\varphi_{\mu \nu} = 0$ \cite{Shabad:1988}. We would like to
emphasize that in contrast to the case of unmagnetized
electron-positron plasma where introducing $u_\mu$ is required to
present the two- and three-photon vertex in a covariant
form~\cite{Fradkin:1965, Weldon:1982}, in the presence of a magnetic
field it is sufficient to take the substitutions
$f_{\pm}(E)~\to~f_{\pm}(\upsilon p)$ in the electron and positron
distribution functions in (\ref{eq:J1}) and (\ref{eq:J2}). The fact
is that an orthogonal basis can be constructed only from the field
tensor and the 4-momentum vector\cite{Shabad:1988}:
\begin{eqnarray}
 b_{\mu}^{(1)} = \frac{(\varphi q)_\mu}{\sqrt{q^2_{\mprp}}}, \qquad
 b_{\mu}^{(2)} = \frac{(\tilde \varphi q)_\mu}{\sqrt{q^2_{\mprl}}},
 \label{eq:b1b2}
\end{eqnarray}
\vspace{-5mm}
\begin{eqnarray}
 b_{\mu}^{(3)} = \frac{q^2 \, (\Lambda q)_\mu - q_\mu \, q^2_{\mbox{\tiny $\bot$}}}
 {\sqrt{q^2 q^2_{\mprl} q^2_{\mprp}}}, \qquad
 b_{\mu}^{(4)} = \frac{q_\mu}{\sqrt{q^2}}.
 \label{eq:b3b4}
\end{eqnarray}
With this basis  any tensor can be represented in the covariant form.

The amplitude of the process written in the form (\ref{eq:M}) can be
used to obtain the axion two-photons ($a \to \gamma \gamma$) and
two-photons two-neutrino ($\gamma \gamma \to \nu \bar \nu$)
interaction amplitude by making the substitutions
%
%\begin{widetext}
\begin{eqnarray}
 \M_{a \to \gamma \gamma} =
\M \Big ( \varepsilon_\mu \to q_\mu \frac{i g_{a e}}{ 2 m e}, \Pi_{\mu \nu \rho} \to \tilde \varphi_{\mu \sigma} \Pi_{\sigma \nu \rho} \Big )
\label{eq:Magg}
\end{eqnarray}
and
\begin{eqnarray}
 \M_{\gamma \gamma \to \nu \nu} &=&
\M \Big (\varepsilon_\mu \to j_\mu \frac{G_F}{\sqrt{2} e}, \Pi_{\mu \nu \rho} \to C_V \Pi_{\mu \nu \rho}
\nonumber \\
&+& C_A \tilde \varphi_{\mu \sigma} \Pi_{\sigma \nu \rho}\Big )
\label{eq:Mggnn}
\end{eqnarray}
%
%\end{widetext}
correspondingly~\cite{RCh05}. Here, $C_V$ and $C_A$ are the
vector and axial-vector constants of the effective $\nu \nu e e$
Lagrangian of the standard model,
$$
C_V = \pm 1/2 + 2 \sin^2 \theta_W, \quad C_A = \pm 1/2,
$$
where $\theta_W$ is the Weinberg angle, the upper sign corresponds  to an
electron neutrino, the lower sign corresponds to muon and tau
neutrinos, $j_\mu$ is the Fourier transform of the neutrino current,
and $g_{ae}$ is the dimensionless axion-electron coupling constant.

%%%%%%%%%%%%%%%%%%%%
\section{Kinematics and selection rules}
\label{Sec:3}

As was mentioned in the Sec. I, the presence of magnetized plasma
influences  not only the process amplitude but the photon dispersion
properties also and consequently it could change the kinematics of the
process.

It  is convenient to
describe the propagation of the photon in any active medium in 
terms of the normal modes (eigenmodes). Then, the
polarization and dispersion properties of the normal modes are connected
with the eigenvectors and the eigenvalues of the polarization operator
correspondingly.

To calculate the polarization operator, it is useful to represent it 
as an expansion over the 
basis (\ref{eq:b1b2}),~(\ref{eq:b3b4})                               
\begin{eqnarray}
{\cal P}_{\mu \nu}(q) = \sum\limits_{i,j = 1}^{4} b^{(i)}_\mu b^{(j)}_\nu
{\cal P}^{(i, j)}(q).
\label{eq:Pbas}
\end{eqnarray}
Note that the terms with $i,j = 4$ may be omitted in
(\ref{eq:Pbas}) due to  the gauge invariance of the polarization
operator  $q_\mu {\cal P}_{\mu \nu }(q) = {\cal P}_{\mu \nu }(q)
q_\nu = 0$. In the magnetic field without plasma  the polarization
operator is known to be diagonal in such basis~\cite{Shabad:1988}.
Therefore one can write ${\cal P}^{(i, j)}(q)$ in the following way:
\begin{eqnarray}
 {\cal P}^{(i, j)}(q) = \delta_{i j} {\cal P}_B^{(j)}(q) + {\cal P}_{pl}^{(i, j)}(q),
\label{eq:Pij}
\end{eqnarray}
where ${\cal P}_B^{(j)}(q)$ is the magnetized vacuum contribution and
${\cal P}_{pl}^{(i, j)}(q)$ is  the contribution arising from the photon
forward scattering on electrons and positrons in plasma.

In the kinematic region where the absorption of the photon is
disregarded, the  polarization operator is the Hermitian matrix
$$
{\cal P}_{\mu \nu}(q) = [{\cal P}_{\nu \mu}(q)]^{*}.
$$
This implies that the matrix ${\cal P}_{pl}^{(i, j)}(q)$ is also Hermitian
$$
{\cal P}_{pl}^{(i, j)}(q) = [{\cal P}_{pl}^{(j, i)}(q)]^{*}
$$
in the same region.

Previously, the polarization operator in the presence of a  magnetic
field was investigated in several papers. In the strongly magnetized
vacuum limit it could be taken e.g. from~\cite{Shabad:1988}:
\begin{eqnarray}
\P_B^{(1)}(q) &\simeq& - \frac{\alpha}{3 \pi}\,
q_{\mbox{\tiny $\bot$}}^2 -
q^2\, \Lambda(B), \label{eq:P1B}
\\[2mm]
{\cal P}_B^{(2)}(q) &\simeq& -\frac{2 eB \alpha}{\pi}
H\left(\frac{q^2_{\mbox{\tiny $\|$}}}{4m^2} \right)
- q^2 \, \Lambda(B), \label{eq:P2B}
\\[2mm]
\P_B^{(3)}(q) &\simeq&  - \, q^2 \, \Lambda(B),
\label{eq:P3B}
\end{eqnarray}
where
$$
\Lambda(B) = \frac{\alpha}{3 \pi}\,\left[1.792 - \ln
(B/B_e)\right].
$$

The polarization operator in magnetized plasma   was studied
in~\cite{Rojas1979,Rojas1982,Shabad:1988}. However,  to verify the
method described in Sec. \ref{Sec:2} we have calculated ${\cal
P}_{pl}^{(i, j)}(q)$ in strongly magnetized  plasma\footnote{Hereafter we consider only the case of the
charge-symmetric ($\mu = 0$) electron-positron plasma. The case of the
nonzero chemical potential will be investigated elsewhere.}:
\begin{eqnarray}
 {\cal P}_{pl}^{(1, 1)}(q) = - \frac{2 \alpha}{\pi eB} \int \frac{d p_z}{E}\,
 (pq)^2_{\mbox{\tiny $\|$}}\,
 [e^{E/T} \,+\, 1]^{-1}, \label{eq:Ppl11}
\end{eqnarray}
\begin{eqnarray}
{\cal P}_{pl}^{(2, 2)}(q) =  - \frac{8 eB \alpha}{\pi}\,
\prll{q}^2 m^2 \int \frac{dp_z}{E} \,
\frac{[e^{E/T} \,+\, 1]^{-1}} {\prll{q}^4 - 4 \prll{(pq)}^2}\, ,
%{\cal J}_1(q_{\mbox{\tiny $\|$}}),
\label{eq:Ppl22}
\end{eqnarray}
\begin{eqnarray}
\label{eq:Ppl23}
{\cal P}_{pl}^{(2, 3)}(q) =  - \frac{2 \alpha}{\pi eB}\,
\sqrt{\frac{q^2_{\mbox{\tiny $\bot$}}}{q^2}} \int \frac{d p_z}{E}\,
(p \tilde \varphi q)\, (pq)_{\mbox{\tiny $\|$}} 
[e^{E/T} \,+\, 1]^{-1}, \nonumber \\
\end{eqnarray}
\begin{eqnarray}
{\cal P}_{pl}^{(3, 3)}(q) = \left(\frac{q^2_{\mbox{\tiny $\bot$}}}{q^2} - 1 \right)\,
{\cal P}_{pl}^{(1, 1)}(q),
\label{eq:Ppl33}
\end{eqnarray}
\begin{eqnarray}
{\cal P}_{pl}^{(1, 2)}(q) &=& {\cal P}_{pl}^{(1, 3)}(q) = 0.
\label{eq:Ppl12_13}
\end{eqnarray}

To obtain the eigenvalues one should diagonalize the matrix ${\cal
P}^{(i,j)}(q)$. It is seen from (\ref{eq:Ppl11})-(\ref{eq:Ppl12_13})
that all of the nonzero components of the matrix ${\cal P}_{pl}^{(i,j)}(q)$
except ${\cal P}_{pl}^{(2,2)}(q)$ have an inverse dependence on
the magnetic field strength. It means that these elements give a small
contribution to diagonalized ${\cal P}^{(i)}(q) \equiv {\cal
P}^{(i,i)}(q)$ in comparison with ${\cal P}_{B}^{(i)}(q)$. Finally
the polarization operator eigenvalues can be written to within
$O(B_e/B)$ in the following form:
\begin{eqnarray}
{\cal P}^{(1)}(q) &\simeq&  {\cal P}_B^{(1)}(q),
\label{eq:P1}
\\[2mm]
{\cal P}^{(2)}(q) &\simeq& {\cal P}_B^{(2)}(q) + {\cal P}_{pl}^{(2,2)}(q),
\label{eq:P2}
\\[2mm]
{\cal P}^{(3)}(q) &\simeq&  {\cal P}_B^{(3)}(q).
\label{eq:P3}
\end{eqnarray}
This result is in agreement with the previous one obtained by  different
methods in Refs.~\cite{Rojas1979,Rojas1982,Shabad:1988}.

\begin{figure}
\centerline{\includegraphics[width = 9cm]{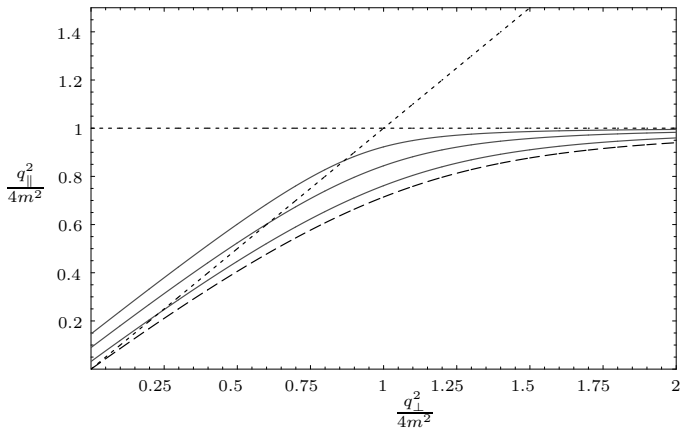}}
\vspace*{-2mm} \caption{Photon dispersion laws in strong magnetic
field $B/B_e = 200$ and neutral plasma vs temperature:  $T = 1$ MeV
(upper curve), $T = 0.5$ MeV (middle curve), $T = 0.25$ MeV (lower
curve). Photon dispersion without plasma is depicted by the dashed line.
The dotted line corresponds to the vacuum dispersion law, $q^2 = 0$. The
angle $(\theta)$ between the photon momentum and the magnetic field direction
is $\pi/2$. } \label{fig:disT}
\end{figure}

The dispersion properties of the eigenmodes now can be found  from
the corresponding equations
\begin{eqnarray}
q^2 - {\cal P}^{(\lm)}(q) = 0 \qquad (\lm = 1, 2, 3).
\label{eq:disper}
\end{eqnarray}
Their analysis shows that the waves with $\lambda = 1, 2$ and
the polarization vectors
\begin{eqnarray}
\varepsilon_\mu^{(1)}(q) = b_\mu^{(1)},
\qquad
\varepsilon_\mu^{(2)}(q) = b_\mu^{(2)}
\label{eq:epsilon}
\end{eqnarray}
are the only physical modes in the case under consideration, just as in
a pure magnetic field\footnote{ Symbols 1 and 2 correspond to the
$\|$ and $\perp$ polarizations in a pure magnetic
field in Adler's notation~\cite{Adler:1971} or $E$- and $O$- modes in charge-symmetric
 magnetized
plasma~\cite{Thompson:1995}.}. The third eigenmode, $\lambda  = 3$,
does not correspond to any real wave~\cite{Shabad:1988}. Indeed,
the substitution of the expression for ${\cal P}^{(3)}(q)$ into
(\ref{eq:disper})  gives the equation that has the only solution
$q^2 = 0$. Then the corresponding eigenvalue $b_\mu^{(3)}$ appears
to be proportional to $q_\mu$ and, therefore, can be eliminated by
a suitable gauge transformation.

\begin{figure}
\centerline{\includegraphics[width=9cm]{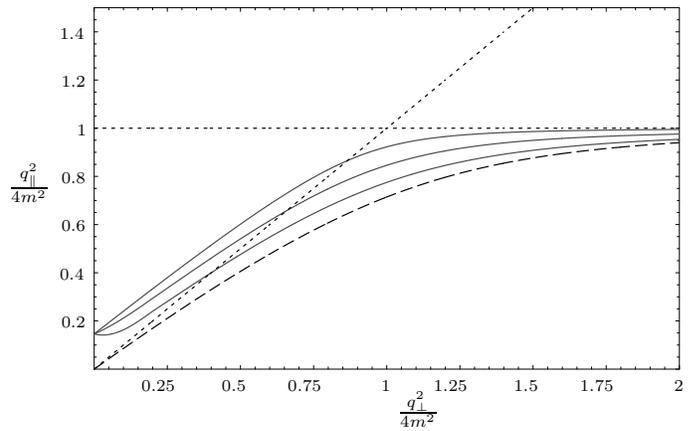}} \vspace*{-2mm}
\caption{Photon dispersion laws in strong magnetic field $B/B_e =
200$ and neutral plasma $(T = 1 \mbox{MeV})$ vs the angle between
magnetic field direction and photon momentum:  $\theta = \pi/2$
 (upper curve), $\theta = \pi/6$  (middle curve), $\theta = \pi/12$
 (lower curve).
Photon dispersion without plasma is depicted by the dashed line. The dotted
line corresponds to the vacuum dispersion law, $q^2 = 0$. }
\label{fig:disTheta}
\end{figure}

Let us consider the kinematics of the process under consideration. In
the magnetized vacuum it is convenient to represent the dispersion
law of the mode $\lambda$ in the form $q_{\mbox{\tiny $\|$}}^2 =
f^{(\lambda)}(q_{\mbox{\tiny $\bot$}}^2)$~\cite{Shabad:1988}. From
the energy conservation law it follows that a photon splitting is
kinematically allowed only if the condition holds:
\begin{eqnarray}
 \sqrt{q_{\mbox{\tiny $\|$}}^2} \ge \sqrt{q'^2_{\mbox{\tiny $\|$}}} +
 \sqrt{q''^2_{\mbox{\tiny $\|$}}}\, .
\label{eq:Kinematic}
\end{eqnarray}
 Using the dispersion law it can be written as
\begin{eqnarray}
 \sqrt{f^{(\lambda)}(q_{\mbox{\tiny $\bot$}}^2}) \ge \sqrt{f^{(\lambda')}
 (q'^2_{\mbox{\tiny $\bot$}})} +
 \sqrt{f^{(\lambda'')}(q''^2_{\mbox{\tiny $\bot$}})}.
\end{eqnarray}
The analysis of the dispersion equation solutions shows that  in the
region below the pair creation threshold ($q^2_{\mbox{\tiny $\|$}}~=~4 m^2$) 
the functions $f^{(\lambda)}(q_{\mbox{\tiny
$\bot$}}^2)$ are the monotonic single-valued  and the inequalities
\begin{eqnarray}
f^{(\lambda)}(q_{\mbox{\tiny $\bot$}}^2) \le q_{\mbox{\tiny $\bot$}}^2,
 \quad f^{(2)}(q_{\mbox{\tiny $\bot$}}^2) < f^{(1)}(q_{\mbox{\tiny $\bot$}}^2)
\label{eq:BfieldEnq}
\end{eqnarray}
hold throughout the interval $0 \le \prp{q}^2 \le \infty $. From
these conditions and the inequality (\ref{eq:Kinematic}), it
immediately follows that only the splitting channels $\gamma_1 \to
\gamma_2 \gamma_2, \gamma_1 \to \gamma_1 \gamma_2$ are kinematically
allowed below the pair creation threshold~\cite{Adler:1971,Shabad:1983, Usov:2002, Chistyakov:1998}. 
It coincide with Adler's selection rules in a weak magnetic field~\cite{Adler:1971}. 

It is usually assumed that  the 
photon splitting can be neglected in the region $q^2_{\mbox{\tiny $\|$}} > 4 m^2$ 
in comparison with the process of the pair creation
$\gamma \to e^{+} e^{-}$. 
This is true in the case of a not too strong
magnetic field, $B \lesssim B_e$, when the resonances of  the
polarization operator corresponding to the creation thresholds of electrons and
positrons on the different Landau levels are close to each other. In a 
strong magnetic field, the gap between the two first resonances becomes wide. In this case
the 2-mode photon can decay into the  $e^{+} e^{-}$-pair just above
the first resonance $q^2_{\mbox{\tiny $\|$}} = 4 m^2$, whereas the
lowest 1-mode pair creation threshold is
$q^2_{\mbox{\tiny $\|$}} = 4 m^2(1/2 +\sqrt{1/2 + B/B_e})^2 \simeq 4m^2 B/B_e$.
Therefore in the kinematical region $4 m^2 < q^2_{\mbox{\tiny $\|$}}< 4 m^2 B/B_e$ the 
only QED absorption
mechanism for the 1-mode is the photon splitting. It is not difficult to understand that the 
1-mode photon
in this region can split by the same channels $\gamma_1 \to
\gamma_2 \gamma_2, \gamma_1 \to \gamma_1 \gamma_2$ if the 2-mode photons in the 
final state are created with $q^2_{\mbox{\tiny $\|$}}$ below $4 m^2$~\cite{Chistyakov:1998}.
Strictly speaking, the 2-mode photon also can split above the pair creation threshold. 
However, as it was mentioned above, the corresponding absorption rates are negligible 
in comparison with the rate of the process $\gamma_2 \to e^{+} e^{-}$.

The presence of plasma could change the selection rules described above. A 
comparison of the Eqs. (\ref{eq:P1B})-(\ref{eq:P3B}) and
(\ref{eq:P1})-(\ref{eq:P3}) shows that   only the
eigenvalue $\P^{(2)}(q)$ is modified in plasma. It means that  the dispersion
law of the mode 1 photon is the same  as in the magnetized vacuum, where
it is spacelike and its deviation from the vacuum law, $q^2 = 0$,
is negligibly small. On the other hand, the dispersion properties
of the mode 2 photon essentially differs from the ones in magnetized vacuum.
In this case the dispersion law in the form of the relation between $q^2_{\mbox{\tiny $\|$}}$ 
and $q^2_{\mbox{\tiny $\bot$}}$ additionally depends on the 
angle between the magnetic field
direction and the photon momentum $q^2_{\mbox{\tiny $\|$}} = f^{(2)}(q^2_{\mbox{\tiny $\bot$}}, \theta)$. 
% $q^2_{\mbox{\tiny $\|$}} = f_{\lambda}(q^2_{\mbox{\tiny $\bot$}})$.

In Figs.~\ref{fig:disT} and~\ref{fig:disTheta},  the photon
dispersion laws are depicted  both in a strong magnetic field and in magnetized plasma at
various temperatures, angles and photon momenta.
One can see that contrary to the pure magnetic field case in plasma
there is the region with $q^2 > 0$  below the pair creation
threshold where the inequalities hold, opposite of the relations
(\ref{eq:BfieldEnq}). It is connected with the appearance  of
the plasma frequency  in the presence of  real electrons and
positrons which can be defined from the equation
\begin{equation}
\omega_{pl}^2 - \P^{(2)}(\omega_{pl}, {\mathbf k} \to 0 ) = 0.
\label{eq:omegapl}
\end{equation}

These facts lead to  new polarization selection rules: in the
region $q^2 > 0$, a new photon splitting channel $\gamma_2 \to
\gamma_1 \gamma_1 $ forbidden  in the magnetic field without plasma
is possible, while the splitting channels $\gamma_1 \to \gamma_2
\gamma_2$ and $\gamma_1 \to \gamma_1 \gamma_2$ allowed in the pure
magnetic field, are forbidden. In the region $q^2 < 0$ polarization
selection rules are the same ones as in a magnetized vacuum. Strictly
speaking,  the dependence of the dispersion law on the angle between
magnetic field direction and photon momentum could lead to the
permission of the additional splitting channel, e.g. $\gamma_2 \to
\gamma_2 \gamma_2$ or $\gamma_2 \to \gamma_1 \gamma_2$. However the
numerical analysis shows that these transitions are forbidden under
considered conditions.

As it follows from~(\ref{eq:P2}), the eigenvalue of the polarization
operator ${\cal P}^{(2)}(q)$ has singular behavior in the vicinity
of the pair-creation threshold:
\begin{eqnarray}
{\cal P}^{(2)}(q) \simeq -\frac{2 \alpha \, eB \,  m}{\sqrt{4 m^2
- q_{\mbox{\tiny $\|$}}^2 }}\, \tanh \frac{\omega}{4 T}.
\end{eqnarray}
This fact leads to the necessity of taking into account of a
wave function renormalization for the photon of mode~2
\beq
\ee_{\alpha}^{(2)}(q) \to \ee_{\alpha}^{(2)}(q) \sqrt{Z_2}, \quad
Z^{-1}_2 = 1 - \frac{\partial {\cal P}^{(2)}(q)}{\partial \omega^2}.
\label{eq:Z}
\eeq

The partial amplitudes for the channels $\gamma_1 \to \gamma_1
\gamma_2$, $\gamma_1 \to \gamma_2 \gamma_2$ and $\gamma_2 \to
\gamma_1 \gamma_1$ could be obtained from (\ref{eq:M}) by
substitution of the corresponding polarization vectors
(\ref{eq:epsilon})
\begin{eqnarray}
\label{eq:Mpartial}
\M_{\lambda \to \lambda' \lambda ''} = \, \ee^{(\lambda)}_\mu(q) 
\ee^{(\lambda')*}_\rho(q') \ee^{(\lambda'')*}_\nu(q'') \, \Pi_{\mu \nu \rho}.
%[\Pi^{(0)}_{\mu \nu \rho} + \Pi^{(1)}_{\mu \nu \rho}],
\end{eqnarray}
\noindent Then, 
\begin{eqnarray}
{\cal M}_{1 \to 1 2} &=& - i 4\pi \left(
\frac{\alpha}{\pi}\right)^{\mbox{\tiny $3/2$}} \frac{(q'\varphi
q'')(q'\tilde \varphi q'')} {[\prp{q'^2} \prll{q''^2}
\prp{q^2}]^{\mbox{\tiny $1/2$}}} {\cal G}(q''_{\mbox{\tiny $\|$}}),
\label{eq:M112}
\\[3mm]
{\cal M}_{1 \to 2 2} &=& - i 4\pi \left(
\frac{\alpha}{\pi}\right)^{\mbox{\tiny $3/2$}} \frac{(q'
q'')_{\mbox{\tiny $\|$}}} {[\prll{q'^2} \prll{q''^2}
q^2_{\mprp}]^{\mbox{\tiny $1/2$}}} \label{eq:M122}
\\[2mm]
&\times& \left \{(q q'')_{\mbox{\tiny $\bot$}} {\cal G}(q'_{\mbox{\tiny $\|$}}) +
(q q')_{\mbox{\tiny $\bot$}} {\cal G}(q''_{\mbox{\tiny $\|$}}) \right \},
\nonumber
\\[3mm]
{\cal M}_{2 \to 1 1} &=& {\cal M}_{1 \to 1 2} (q \leftrightarrow q''), \quad
\label{eq:M211}
\end{eqnarray}
where ${\cal G}(\prll{q})$ is given by Eq. (\ref{eq:Gq}).

\section{Photon absorption rate}
\label{Sec:4}

%\subsection{Photon Splitting}
%\label{Subsec:splitting}

To analyze the efficiency of the process under consideration and to
compare it with other competitive reactions we  calculate the photon
absorption rate due to photon splitting which can be defined in the
following way:

\begin{eqnarray}
&&W_{\lambda \to \lambda^{'} \lambda^{''}} = \frac{g_{\lm'
\lm''}}{32 \pi^2 \omega_{\lambda}} \int \left \vert {\cal
M}_{\lambda \to \lambda^{'} \lambda^{''}} \right \vert^2
Z_{\lambda}Z_{\lambda^{'}}Z_{\lambda^{''}}
\nonumber \\[2mm]
&&\times (1 + f_{\omega'})(1 + f_{\omega''})
\nonumber \\
&&\times \delta (\omega_{\lambda}({\bf k}) -
\omega_{\lambda^{'}}({\bf k} - {\bf k^{''}}) -
\omega_{\lambda^{''}}({\bf k^{''}})) \frac{d^3
k^{''}}{\omega_{\lambda^{'}} \omega_{\lambda^{''}}},
\label{eq:Wsplitting}
\end{eqnarray}

\noindent where $f_{\omega} = [e^{\omega/T} - 1]^{-1}$ is the
photons distribution function and the factor $g_{\lm' \lm''} = 1 -
(1/2)\,\delta_{\lambda' \lambda''}$ is inserted to account for the
possible identity of the final photons.
\begin{figure}
\includegraphics[width=8cm]{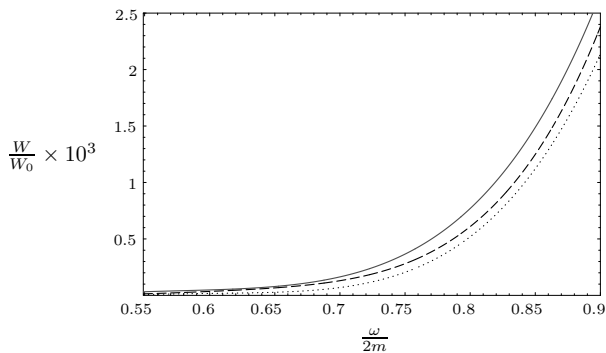} \vspace*{-2mm}
\caption{ Absorption rate of the channel $\gamma_1 \to
\gamma_1 \gamma_2$ in  strong magnetic field ($B/B_e = 100$)  at
temperatures  50 keV (solid line) and 250 keV (dotted line). The
dashed curve represents the probability in a pure magnetic field ($T
= \mu = 0$)  \cite{Chistyakov:1998}. $W_0 \, = \, (\alpha/\pi)^3\, m
\simeq 3.25\cdot 10^2 cm^{-1}$and  $\theta = \pi/2$. \label{fig:W112_low}}
\end{figure}

\subsection{Channels $\gamma_1 \to \gamma_1 \gamma_2$ and $\gamma_1 \to \gamma_2 \gamma_2$}
\label{SubSec:4.1}

In general case the reaction rates~(\ref{eq:Wsplitting}) can be calculated only 
numerically. However in some limiting cases it is possible to obtain the simple
expression for the rates. Analysis shows that  in the low-temperature ($T \ll m$) and low-energy ($\omega \ll m$) limits the
modification of the amplitudes (\ref{eq:M112})-(\ref{eq:M211}) in
the presence of plasma is  small in comparison with the expressions
in pure magnetic field. The main manifestation of plasma influence
arises from taking account of 2-mode photon dispersion in energy
conservation law. In the limit under consideration it is convenient
to use the following approximation for photon dispersion law:

\begin{eqnarray}
q^2_{\mbox{\tiny $\|$}} \simeq \omega^2_{pl} \sin^2{\theta} + (1 - \xi)
q^2_{\mbox{\tiny $\bot$}}, %
\label{eq:disp_low}
\end{eqnarray}
where
\begin{eqnarray}
\omega^2_{pl} &=& \frac{4 \alpha \pi}{m} n_e ,
\label{eq:q_disp}\\
n_e &\simeq& eB\sqrt{\frac{mT}{2\pi^3}}\,e^{-m/T}. \label{eq:ne}
\end{eqnarray}
%$$
%q^2_{pl} = \frac{2 \alpha eB}{\pi} e^{-m/T} \sqrt{\frac{2 \pi
%T}{m}} \sin^2 \theta
%$$
%
\begin{figure}
\includegraphics[width=7.7 cm]{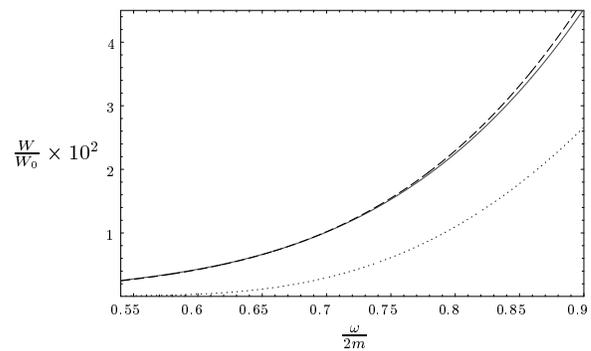} \vspace*{-2mm}
\caption{ Absorption rate of the channel $\gamma_1 \to
\gamma_2 \gamma_2$ for the same parameters and notation as those
in Fig.4. \label{fig:W122_low}}
\end{figure}

\begin{figure*}
\includegraphics{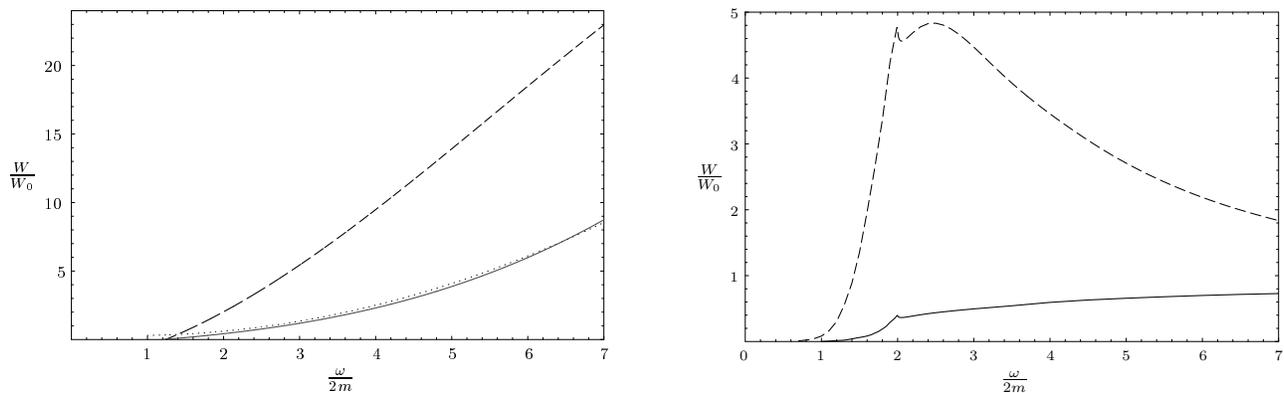} \vspace*{-2mm}
\caption{The dependence of the absorption rates of channel $\gamma_1 \to
\gamma_1 \gamma_2$ (left panel) and $\gamma_1 \to \gamma_2
\gamma_2$ (right panel) on energy in a strong magnetic field $B/B_e
\, = \, 200$ and neutral
 $(T \, = \, 1 \mbox{MeV})$ plasma.
The dashed line corresponds to the probability in the pure magnetic
field $(T \, = \, 0)$~\cite{Chistyakov:1998}. The asymptote
(\ref{eq:W112HT}) is depicted by dots. Here, $\theta = \pi/2$. 
%Here  $W_0 \, = \,(\alpha/\pi)^3\, m$.
\label{fig:W112_W122_high}}
\end{figure*}

\noindent Here $n_e$ is the number of electron and positron density in
strongly magnetized, charge-symmetric low-temperature plasma;
$\theta$ is the angle between the  photon momentum ${\bf k}$ and the
magnetic field direction ${\bf B}$;
%(note that $q^2_{pl}$ in the limit ${\bf k} \to 0$ is reduced to $\omega_{pl}^2$ );
$\xi =
\frac{\alpha}{3 \pi} \frac{B}{B_e}$ is the parameter characterizing
magnetic field influence (it is assumed small in the limit under
consideration).

\noindent Then, using the photon dispersion law (\ref{eq:disp_low}) one
can obtain the following expressions of photon splitting rates in a 
low-temperature case:

\begin{eqnarray}
W_{1 \to 12} &\simeq& \frac{\alpha^3 \xi^2 m}{288 \pi^2} \, \left (
\frac{T}{m} \right )^5 \, {\cal F}_1
\left(\frac{\omega}{T},\sqrt{1-u^2}
\right),%
\label{eq:W112LT}\\
W_{1 \to 22} &\simeq& \frac{\alpha^3  m}{72 \pi^2} \, \left (
\frac{T}{m} \right )^5 \, {\cal F}_2
\left(\frac{\omega}{T},\sqrt{1-u^2} \right),%
\label{eq:W122LT}
\end{eqnarray}
where $u \, = \, \cos{\theta}$ and ${\cal F}_1(y,z)$ and ${\cal
F}_2(y,z)$ are the integrals
%\begin{widetext}
\begin{eqnarray}
{\cal F}_1(y,z) &=& \frac{1}{y^2 z} \, \int \limits_{\delta}^{y z}
\frac{dx}{x^2} \, \frac{[x^2 - \delta^2]^4}{[1 - \exp(-x)]}
\nonumber \\
&\times& \frac{1}{[1 -
\exp(x - y)(1 - \frac{\xi}{2 y}(x^2 - \delta^2))]},%
\label{eq:F1}\\[3mm]
{\cal F}_2(y, z) &=& z \, \Theta(y z - 2 \delta) \int
\limits_{\delta}^{\lambda(y,z) } dx \frac{[ x (y z - x) -
\delta^2]^2}{[1 - \exp(-x)]}
\nonumber \\
&\times& \frac{1}{[1 - \exp(x - y)(1 - \frac{\xi}{2 y}(x^2 -
\delta^2))]}, \label{eq:F2}
%\lambda(y,z) &=& \frac{y z}{2} \left( 1+\sqrt{1-\frac{4
%\delta^2}{y^2 z^2}}\right)\\
\end{eqnarray}
%\end{widetext}
%
where
$$
\lambda(y,z) = \frac{y z}{2} + \sqrt{\frac{y^2 z^2}{4}-\delta^2}
$$
and $\delta \equiv (\omega_{pl}/T) \xi^{-1/2}$; $\Theta(x)$ is theta function.
Note, that in the limit
$T \to 0$ one can obtain the expressions of photon splitting rates
in strongly magnetized vacuum:
\begin{eqnarray}
W_{1 \to 12} &\simeq& \frac{\alpha^3 \xi^2 m}{2016 \pi^2} \left (
\frac{\omega}{m} \right )^5 \sin^6 \theta, \label{eq:W112LB}\\
W_{1 \to 22} &\simeq& \frac{\alpha^3 m}{2160 \pi^2} \left (
\frac{\omega}{m} \right )^5 \sin^6 \theta, \label{eq:W122LB}
\end{eqnarray}
The expression (\ref{eq:W122LT}) may be easily derived from Eq.
(23) of \cite{Adler:1971} (see also, e.g., \cite{Thompson:1995},
\cite{HBG:1997}). To our knowledge the expression of the photon
splitting rate for channel $\gamma_1 \to \gamma_1 \gamma_2$ in low-energy limit was not published before.

The comparison of photon splitting rates in strongly magnetized
plasma  (\ref{eq:W112LT}), (\ref{eq:W122LT}) and vacuum
(\ref{eq:W112LB}), (\ref{eq:W122LB}) shows that electron-positron
background and thermal photons  have an opposite influence on the
process under consideration. On one hand, the presence of  $e^+ e^-$
plasma leads to the decreasing of the amplitude and phase space of
the reactions. On the other hand photon splitting rates increase due
to bosonic final state stimulation effect. From the analysis of the
result obtained (\ref{eq:W112LT}), (\ref{eq:W122LT}) follows that in
the low-energy and low-temperature limit the plasma influence leads
to the increase of the photon splitting rate beside the magnetized
vacuum case\footnote{The same is true for a wider range of
energies, see the solid line in Figs.~\ref{fig:W112_low} and
\ref{fig:W122_low}.}. However, numerical calculations show
 that in the range $\omega \sin{\theta}\le 2 m$ rates decrease with
 a temperature increase and become smaller than those in pure
 magnetic field at some values of temperature
 (see the dotted line in Figs.~\ref{fig:W112_low} and
\ref{fig:W122_low}).

\begin{figure*}
\includegraphics{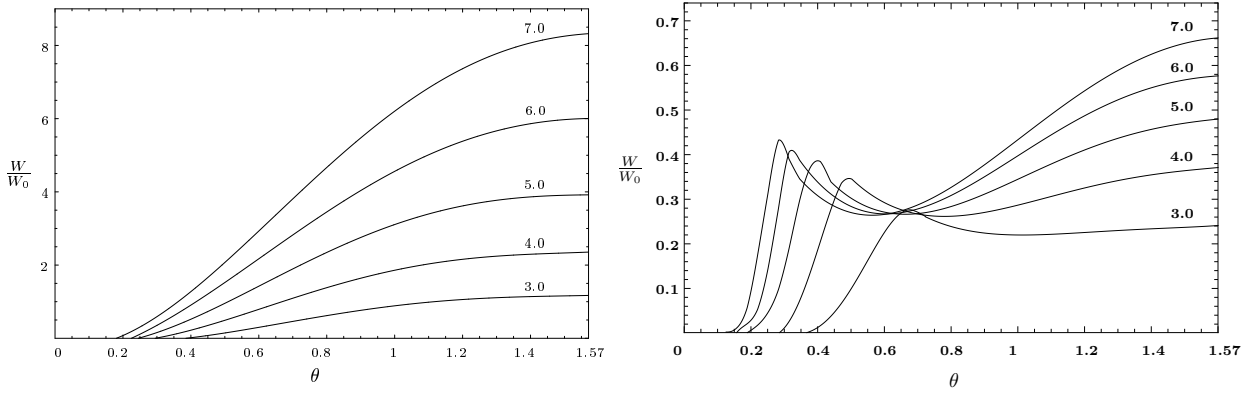}\vspace*{-2mm} \caption{The
dependence of the absorption rates of the channels $\gamma_1 \to
\gamma_1 \gamma_2$ (left panel) and $\gamma_1 \to \gamma_2 \gamma_2$
(right panel) on an angle
between initial photon momentum and magnetic field direction at
different initial photon energies ($B/B_e \, = \, 200, T \, = \, 1
\mbox{MeV}$). The numbers above the curves correspond to the values of
the ratio $\omega / 2 m$.
 \label{fig:W112ang}}
\end{figure*}

The results obtained show that the rate $W_{1 \to 1 2}$ is
considerably suppressed compared to $W_{1 \to 2 2}$. The reason
for this is the same as in the pure magnetic field case. In the
energy region $\omega \sin{\theta} \le 2 m$ the kinematics of the
processes under consideration is closed to the collinear
one\footnote{Strictly speaking, it depends on the value of the
parameter $\xi$. In the presence of the very strong magnetic field
when $\xi \gtrsim 1$ ($B \gtrsim 10^3 B_e$ ), the kinematics of the
photon splitting significantly deviate from the collinear one even
at $\omega \ll m$ .}. In  this case, it is easy to show that in contrast to the channel
$\gamma_1 \to \gamma_2 \gamma_2$ the amplitude $\M_{112}$ (\ref{eq:M112}) contains
terms $(q' \varphi q'') \sim (q' \tilde \varphi q'') \sim \Delta \phi$, where
$\Delta \phi$ is the angular
separation between the two photons in the final state. At low-photon energies
$\Delta \phi \sim \sqrt{\xi}\sin{\theta}$ and
$W_{1 \to 1 2}$ is suppressed by the factor $\xi^2 \sin^2{\theta} \ll 1$ in comparison
with $W_{1
\to 2 2}$.

The situation changes dramatically when the energy of the initial
photons becomes larger than  $2 m/\sin{\theta}$. Let us consider the limit $m^2
\ll \omega^2 \, \sin^2{\theta} \le eB$. The analysis shows that in
this case the main contribution to the probability of the processes
comes from the kinematical region in close vicinity to the pair
creation threshold of the 2-mode photon because the amplitudes
(\ref{eq:M112}) and (\ref{eq:M122}) have the square root singular
behavior in this region. As a consequence, the corresponding
splitting rates might contain the pole singularity. However, taking
account of the photon wave function renormalization corrects the
situation. Indeed, in the limit $q^2_{\mbox{\tiny $\|$}} \to 4 m^2$
it can be shown that the production of the singular function ${\cal
G}(q_{\mbox{\tiny $\|$}})$ in the amplitudes (\ref{eq:M112}),
(\ref{eq:M122}) and the square root of the renormalization function
$Z_2$ is the regular function:
$$
{\cal G}(q_{\mbox{\tiny $\|$}})\sqrt{Z_2} \simeq \frac{ \sqrt{2 }\pi
m}{\sqrt{q^2_{\mbox{\tiny $\bot$}}}}\, \tanh{\frac{\omega}{4 T}}.
$$
Note that in this region the 2-mode photon dispersion law is
simplified and can be defined by the following relationship:
\begin{eqnarray}
\sqrt{1- \frac{q^2_{\mbox{\tiny $\|$}}}{4 m^2}} \simeq \alpha \, \frac{eB}
{q^2_{\mbox{\tiny $\bot$}}} \, \tanh{\frac{\omega}{4 T}}
\label{eq:dispersion_near2m}
\end{eqnarray}

Taking in mind  these facts, we obtain the following
approximation of the photon splitting rates:
\begin{eqnarray}
W_{1 \to 12} &\simeq& \frac{\alpha^3 T^2}{4\omega (1 - u^2)}
\label{eq:W112HT}
\\[3mm]
&\times&\left [(1 \, - \, u)^2 \, {\cal F}_3 \left (\frac{\omega
(1+u)}{2T} \right ) \, + \, (u \to -u) \right ], \nonumber %\\[4mm]
\end{eqnarray}
\begin{eqnarray}
W_{1\to 22} &\simeq& \frac{\alpha^3 m^2}{4\omega} \, \frac{1}{1 \,
-\, \exp{\left[-\frac{\omega}{T}(1 \, - \, u)\right]}}
\label{eq:W122HT} \\[3mm]
&\times& \frac{1}{1 \, - \, \exp{\left[-\frac{\omega}{T}(1 \, + \,
u)\right]}} \nonumber
\\[3mm]
&\times& \left \{\tanh^2\left [ \frac{\omega}{8T}(1 \, - \, u)\right
] \, + \, (u \to -u) \right \},\nonumber
\end{eqnarray}
\noindent where
\begin{eqnarray}
{\cal F}_3(z) \, = \, \int \limits_0^z  \frac{x \tanh^2(x/4) \, dx}
{[1 \, - \, \exp{(-x)}] \, [1 \, - \, \exp{(x \, - \, \omega/T)}]},
\end{eqnarray}

Again, in the limit $T \to 0$ these formulas turn into
the expressions of the photon splitting probabilities in strong magnetic
field in the absence of plasma~\cite{Chistyakov:1998}:
\begin{eqnarray}
W_{1 \to 12} &\simeq& \frac{\alpha^3 \omega \sin^2{\theta}}{16},
\label{eq:W112HB}\\
W_{1 \to 22} &\simeq& \frac{\alpha^3 m^2}{2 \omega},
\label{eq:W122HB}
\end{eqnarray}
from which follow that at high-photon energies the rate of the
channel $\gamma_1 \to \gamma_1 \gamma_2$ dominates over $W_{1 \to 2
2}$. The analysis of the approximations  (\ref{eq:W112HT}),
(\ref{eq:W122HT}) and numerical calculations (see
Fig.~\ref{fig:W112_W122_high}) show that the same relation between
channel rates is kept also in the presence of plasma. One can see
also that photon splitting rates in plasma is suppressed in
comparison with pure magnetic field results at high energies of the
splitting photon.

In addition to the energy dependence of photon splitting
probability it is interesting also to consider the dependence on
angle between initial photon momentum and magnetic field
direction. It is important, e.g., in the problem of radiation
transfer in strongly magnetized plasma (see Sec.~\ref{Sec:6}).
The results of the numerical calculation are depicted in
Fig.~\ref{fig:W112ang} for  different energies of the initial
photon. It is interesting to note that rate of the channel $\gamma_1
\to \gamma_1 \gamma_2$ attains its maximum at $\theta = \pi /2$
while $W_{1 \to 22}$ may have global maximum at $\theta < \pi /2$.

\subsection{Channel $\gamma_2 \to \gamma_1 \gamma_1$}
\label{SubSec:4.2}

\begin{figure}
\centerline{\includegraphics[width = 8 cm]{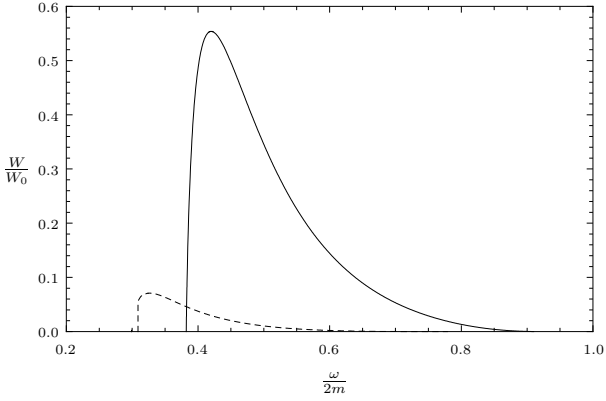}}
\vspace*{-2mm} 
\caption{Absorption
rate of the channel $\gamma_2 \to
\gamma_1 \gamma_1$ in a strong magnetic field ($B/B_e = 200$)  at
temperatures of 1 MeV (solid line) and 500 keV (dashed line).
Here, $\theta = \pi/2$.
\label{fig:W211}}
\end{figure}

As it was found in Sec. \ref{Sec:3} in the presence of strongly
magnetized plasma the "new" photon splitting channel $\gamma_2 \to \gamma_1 \gamma_1$
forbidden in the pure magnetic field, becomes open.
 According to  (\ref{eq:M211}) and (\ref{eq:Wsplitting}) the absorption rate of the  channel
$\gamma_2 \to \gamma_1 \gamma_1$
can be presented  as
\begin{eqnarray}
\label{eq:W211}
W_{2 \to 11} = \frac{\alpha^3}{8\pi^2}\, Z_2\,  {\cal
G}^2(q_{\mprl}) \, \frac{q_{\mprp}^2}{\omega} \, {\cal F}_4\left
(\sqrt{\frac{q_{\mprl}^2}{q_{\mprp}^2}} \right) \, \Theta (q^2)\, ,
\end{eqnarray}
%
%\begin{eqnarray}
%\nonumber
%&&W_{2 \to 11} \simeq \frac{4 \alpha^3}{\pi^3}\,m^3\, Z_2\,  {\cal
%G}^2(q_{\mprl}) \, \frac{q_{\mprp}^2}{\sqrt{q_{\mprl}^2}\, \omega} \, \int^{x_+}_{x_-} dx
%\int^{y_0}_0 dy \times
%\\  [0.1in]
%&&\frac{y^2 \sqrt{y_0^2 - y^2}}
%{(x^2+y^2)[(\sqrt{q_{\mprp}^2} -2m x)^2 + 4m^2 y^2]}\times
%\label{eq:W211}
%\\ [0.1in]
%&&\big \{(1+f_{\omega_+})(1+f_{\omega - \omega_+}) + (\omega_+ \to \omega_-)\big \}
%\Theta (q^2)\, ,
%\nonumber
%\end{eqnarray}

\noindent where
\begin{widetext}
\begin{eqnarray}
\nonumber
{\cal F}_4 (z) &=& \frac{4 z^3}{\pi} \, \int^{1+z}_{1-z} dx \int^{y_0}_0 dy
\frac{y^2 \sqrt{y_0^2 - y^2}}
{(x^2+z^2 y^2)[(2 - x)^2 + z^2 y^2]}\big \{(1+f_{\omega_+})(1+f_{\omega - \omega_+}) + (\omega_+ \to \omega_-)\big \}\, ,
\\
\nonumber
\omega_{\pm} &=& \frac{1}{2 z^2} \, \bigg [\omega (z^2 - 1 + x)
\pm  q_z  z^2 \,\sqrt{y_0^2 - y^2}\, \bigg] \, ,
y_0 =\frac{\sqrt{z^2-1}}{z^2} \, \sqrt{z^2 - (1-x)^2} \, ,
\quad z =  \sqrt{\frac{q_{\mprl}^2}{q_{\mprp}^2}} \,.
\end{eqnarray}
\end{widetext}
 If one can neglect the effect of final photons stimulation emission  $(f_{\omega'} \, = \, 
f_{\omega''}
\, = \, 0)$ the expression for the  function ${\cal F}_4(z)$ is considerably simplified:
$${\cal F}_4(z) \, = \, 2 \ln{z} \, - \, 1 \, + \, z^{-2} \, .$$

In Fig. \ref{fig:W211} the absorption rate of the channel $\gamma_2 \to \gamma_1 \gamma_1$ as a
function of the initial photon energy is depicted for the case of the photon propagation across 
magnetic field direction  at
temperatures of 1 MeV and 500 keV.
One can see that in contrast to the $\gamma_1 \to \gamma_1 \gamma_2$ and 
$\gamma_1 \to \gamma_2 \gamma_2$ channels behavior $\gamma_2 \to \gamma_1 \gamma_1$ the splitting rate decreases rapidly with decreasing temperature.
This is due to an increase of the kinematically allowed region ($q^2 > 0$) for the 
channel under consideration leading to the growth of the process phase space volume 
with temperature increasing.

\begin{figure}
\centerline{\includegraphics[width = 8.1cm]{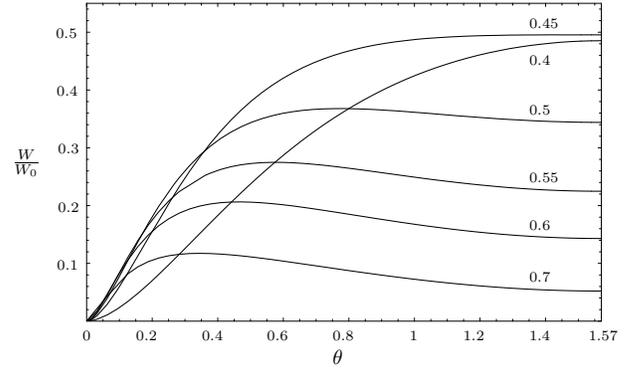}}
\vspace*{-2mm} \caption{The
dependence of the absorption rate of the channel $\gamma_2 \to
\gamma_1 \gamma_1$ on an angle
between initial photon momentum and magnetic field direction at
different initial photon energies ($B/B_e \, = \, 200, T \, = \, 1
\mbox{MeV}$). The numbers above the curves correspond to the values of
the ratio $\omega / 2 m$. \label{fig:W211ang}}
\end{figure}

The dependence of the $\gamma_2 \to \gamma_1 \gamma_1$ splitting rate on an angle between initial photon momentum and magnetic field direction at temperature 1 MeV is depicted in Fig. \ref{fig:W211ang}.  Note that
in the range $\pi/4 \lesssim \theta \le \pi/2$,   $W_{2 \to 11}$ weakly depends on the angle.

\subsection{Photon merging}
\label{Subsec:merging}

Due to the finite photon density in electron-positron plasma the inverse 
process of photon merging $\gamma \gamma \to \gamma$ should be taking into account, e.g., 
in radiation transfer problem~\cite{Thompson:1995}. One can define the absorption rate 
of the process under consideration in the same manner as for the photon splitting:
\begin{eqnarray}
&&W_{\lambda  \lambda^{'} \to \lambda^{''}} = \frac{1}{32 \pi^2
\omega_{\lambda}} \int \left \vert {\cal M}_{\lambda  \lambda^{'}
\to \lambda^{''}} \right \vert^2
Z_{\lambda}Z_{\lambda^{'}}Z_{\lambda^{''}}
\nonumber \\[2mm]
&&\times f_{\omega'}(1 + f_{\omega''})
\nonumber \\
&&\times \delta (\omega_{\lambda}({\bf k}) +
\omega_{\lambda^{'}}({\bf k} + {\bf k^{''}}) -
\omega_{\lambda^{''}}({\bf k^{''}})) \frac{d^3
k^{''}}{\omega_{\lambda^{'}} \omega_{\lambda^{''}}}, \label{eq:WW}
\end{eqnarray}
where amplitudes ${\cal M}_{\lambda  \lambda^{'} \to \lambda^{''}}$
can be obtained from (\ref{eq:M112})-(\ref{eq:M211}) using
crossing symmetry.  We have made  numerical analysis of the
photon merging rate for kinematically allowed channels
in the case of the
initial photon propagation across  magnetic field direction.
The results are presented on Figs. \ref{fig:spmerg_50_250KeV}
and \ref{fig:spmerg_250KeV_1MeV.eps}. As can be seen from these figures, the 
contribution of the channel
$\gamma_1 \gamma_2 \to \gamma_1$ to the  absorption rate
is negligible at low temperature, however it dominates in more hot plasma
(see Fig. \ref{fig:spmerg_250KeV_1MeV.eps}). On the other hand, the channel
$\gamma_2 \gamma_2 \to \gamma_1$ leading at the temperature $T=50$ keV
is kinematically suppressed  at higher temperatures.  The photon splitting
channels are  also suppressed in  hot plasma in the region   $\omega \lesssim 2m$.
It is interesting to note that $W_{2 1 \to 1}$
has resonant amplification  in the vicinity of the  pair creation threshold,  $\omega \simeq 2m$.

We can see now that in
the presence of hot plasma the process of photon merging could
be an effective particle attenuation mechanism, whereas at relatively low temperatures
($T \simeq 50$ keV)  both photon splitting and photon merging processes
play the main role  in the  change  of photon numbers.

\section{Compton scattering}
\label{Sec:5}

Compton scattering is the other type of photon reaction competing  with
photon splitting (merging). The process of photon scattering in
a strongly magnetized plasma, when the initial photon propagates across magnetic
field,  at the different temperatures and   taking into account the photon
dispersion and wave function renormalization was investigated
recently  (see~\cite{RCh09} and references therein). 
Here, we generalize  our results to the case of
 arbitrary angle between initial photon momenta and magnetic field
 direction.

To calculate the amplitude of the process in a strong magnetic field one
should use the Dirac equation solutions at the ground Landau level (\ref{eq:Dirac_solution}).
However, for virtual electron
it is necessary to use the exact propagator,  e.g., in the form of the Landau
level decomposition~\cite{Chodos:1990,KuznOkrug:2011}. As a result we obtain
\begin{eqnarray}
\label{eq:Mcompt}
&&{\cal M}_{\lambda \to \lambda'} = -4\pi \alpha
\exp \left [-\frac{q_{\mprp}^2+\prp{q}'^2 + 2i(q \varphi q')}{4eB} \right ]
\\
\nonumber
&&\times \sum_{n=0}^{\infty}\,
\frac{\ee^{*(\lambda')}_{\alpha}(q')\ee^{(\lambda)}_{\beta}(q)\,T_{\alpha \beta}^n}
{q_{\mprl}^2 + 2(pq)_{\mprl} -2eBn}  + (q \leftrightarrow -q'),
\end{eqnarray}

\noindent where
\begin{eqnarray}
\nonumber
&&T_{\alpha \beta}^n = \frac{2m}{\sqrt{-Q_{\mprl}^2}}\big \{S_n[-(q\tilde \Lambda)_{\alpha}
(Q\tilde \varphi)_{\beta} - (q'\tilde \Lambda)_{\beta}(Q\tilde \varphi)_{\alpha}
\\
[0.1in]
\nonumber
&& + (q \tilde \varphi q')\tilde \Lambda_{\alpha \beta} +
\varkappa (Q\tilde \varphi)_{\alpha}(Q\tilde \varphi)_{\beta}]
\\
[0.1in]
\nonumber
&& - S_{n-1}[(q \tilde \varphi q')(\Lambda_{\alpha \beta} + i\varphi_{\alpha \beta}) -
(q \Lambda)_{\alpha}(Q\tilde \varphi)_{\beta}
\\
[0.1in]
\nonumber
&& - (q' \Lambda)_{\beta}(Q\tilde \varphi)_{\alpha} -
i(Q\tilde \varphi)_{\alpha}(q' \varphi)_{\beta} +
i(Q\tilde \varphi)_{\beta}(q \varphi)_{\alpha}]\big \},
%\end{eqnarray}
\\
[0.1in]
%\begin{eqnarray}
&& S_n = \frac{1}{n!}\left (\frac{(q\Lambda q')-i(q\varphi q')}{2eB} \right )^n.
\end{eqnarray}

\noindent Then substituting the polarization
vectors~(\ref{eq:epsilon}) in~(\ref{eq:Mcompt}) the strong field limit 
one can obtain partial amplitudes for
different polarization configuration of the initial and final
photons in the covariant form~\cite{RCh09}:

\begin{figure}
\includegraphics[width=7cm]{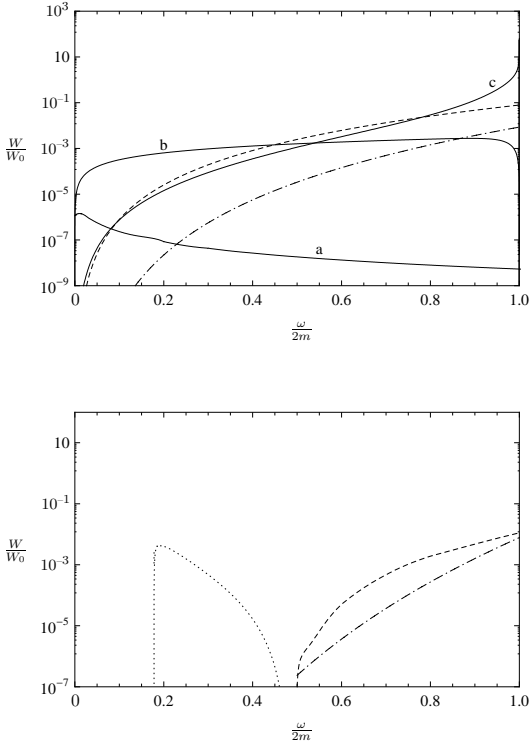}\vspace*{-2mm} \caption{
The dependence of the photon absorption rate
in  strong magnetic field ($B/B_e = 200$)
at temperatures  50 keV (upper panel) and 250 keV
(lower panel) for  channels
$\gamma_2 \to \gamma_1 \gamma_1$ -- dotted line,
$\gamma_1 \to \gamma_2 \gamma_2$ -- dashed-dotted line, $\gamma_1 \to
\gamma_1 \gamma_2$ -- dashed line $\gamma_1 \gamma_2 \to \gamma_1$ --
${\it a}$, $\gamma_2 \gamma_2 \to \gamma_1$ -- ${\it b}$, $\gamma_2 \gamma_1 \to
\gamma_1$ -- ${\it c}$ on energy of the initial
photon. Here, $\theta = \pi/2$.\label{fig:spmerg_50_250KeV}}
\end{figure}

\begin{figure}
\includegraphics[width=7cm]{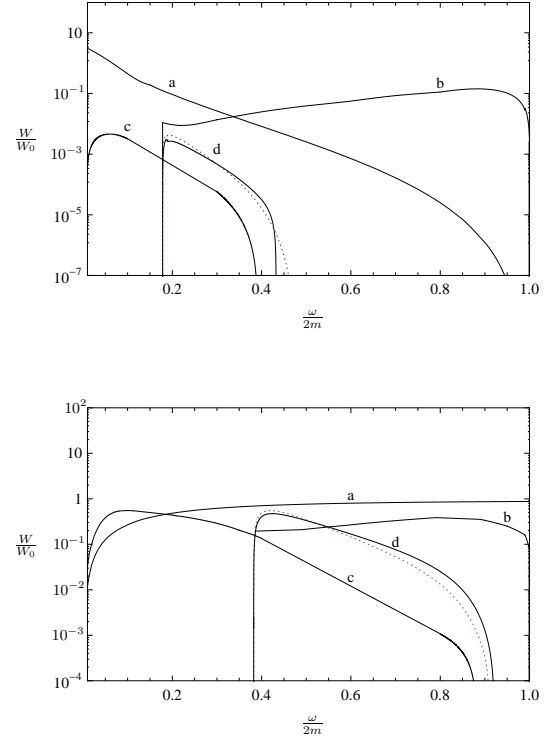}\vspace*{-2mm} \caption{
The dependence of the photon absorption rate
in  strong magnetic field ($B/B_e = 200$)
at temperatures  250 keV (upper panel) and 1 MeV
(lower panel) for  channels
$\gamma_2 \to \gamma_1 \gamma_1$ -- dotted line,
$\gamma_1 \gamma_2 \to \gamma_1$ --
${\it a}$, $\gamma_2 \gamma_2 \to \gamma_1$ -- ${\it b}$, $\gamma_1 \gamma_1 \to
\gamma_2$ -- ${\it c}$, $\gamma_2 \gamma_1 \to
\gamma_1$  -- ${\it d}$ on energy of the initial photon. 
Here, $\theta = \pi/2$. \label{fig:spmerg_250KeV_1MeV.eps}}
\end{figure}
\begin{eqnarray}
{\cal M}_{1 \to 1} = \frac{8 i \pi \alpha m}{eB}\, \frac{(q \varphi
q')(q \tilde \varphi q')} {\sqrt{q^2_{\mprp} q'^2_{\mprp}
(-Q^2_{\mprl})}}, \label{eq:M11}
\end{eqnarray}

\begin{eqnarray}
{\cal M}_{1 \to 2} = \frac{8 i \pi \alpha m}{eB}\, \frac{(q \Lambda
q')(q' \tilde \Lambda Q)} {\sqrt{q^2_{\mprp} q'^2_{\mprl}
(-Q^2_{\mprl})}}, \label{eq:M12}
\end{eqnarray}

\begin{eqnarray}
{\cal M}_{2 \to 1} = -\frac{8 i \pi \alpha m}{eB}\, \frac{(q \Lambda
q')(q \tilde \Lambda Q)} {\sqrt{q^2_{\mprl} q'^2_{\mprp}
(-Q^2_{\mprl})}}, \label{eq:M21}
\end{eqnarray}

\begin{eqnarray}
\label{eq:M22}
&&{\cal M}_{2 \to 2} = 16  \pi \alpha m\, \frac{\sqrt{q^2_{\mprl}
q'^2_{\mprl}}\,\sqrt{(-Q^2_{\mprl})}\, \varkappa} {(q \tilde \Lambda
q')^2 - \varkappa^2 (q \tilde \varphi q')^2}
\\ [2mm]
\nonumber &&\times \left \{1 - \frac{q^2_{\mprp} + q'^2_{\mprp}}{4
eB} + i \frac{(q \varphi q')(q \tilde \varphi q')}{2 eB \varkappa
q^2_{\mprl} q'^2_{\mprl}Q^2_{\mprl}} \right.
\\ [2mm]
\nonumber && \times \left.
 \left [4 m^2 (q \tilde \Lambda q')
 + (q \tilde \Lambda q')^2- \varkappa^2 (q
\tilde \varphi q')^2 \right ]\right \}, 
\end{eqnarray}
%
%\vspace{3mm}

\noindent where $\varkappa = \sqrt{1 - 4m^2/Q^2_{\mprl}}$ and
$Q^2_{\mprl} = (q - q')^2_{\mprl} <0$, $q_{\alpha} = (\omega,{\bf
k})$ and $q'_{\alpha} = (\omega',{\bf k'})$ are the four-momentum of
the initial and final photons correspondingly. 
From the last equation we can see that in the case, when the initial photon propagates
across magnetic field direction, all amplitudes  except ${\cal M}_{2 \to 2}$
are suppressed by magnetic field strength. Therefore one could
expect that mode 2 has the largest scattering absorption rate in
this case.  In turn, in the case when the initial photon propagates almost along ${\bold B}$, 
the amplitude ${\cal M}_{2\to 2}$ is 
suppressed by a small angle between photon momenta and magnetic field direction 
and becomes comparable to other amplitudes.

The general expression of the  photon absorption rates 
is given by the formula~\cite{RCh09}:
\begin{eqnarray}
&&W_{\lambda e^{\pm} \to \lambda^{'} e^{\pm}} = \frac{eB}{16 (2\pi)^4 \omega_{\lambda}}
\int \mid {\cal M_{\lambda \to \lambda^{'} }}\mid^2
Z_{\lambda}Z_{\lambda^{'}}
\nonumber
\\
&&\times f_{\pm}(E)\, (1-f_{\pm}(E')) (1 + f_{\omega'})
\label{eq:Wscatt}\\
&&\times \delta (\omega_{\lambda}({\bf k}) + E - \omega_{\lambda^{'}}({\bf k'}) - E')
\frac{dp_z\,d^3 k^{'}}{ E E' \omega_{\lambda^{'}}},
\nonumber
\end{eqnarray}
where $E = \sqrt{p_z^2 + m^2}$ and $E' = \sqrt{(p_z + k_z - k'_z)^2
+ m^2}$ are the energies of the initial and final electrons
(positrons) correspondingly. In the case of the low-temperature limit ($T
\ll m$) and neglecting the final photon distribution function $(f_{\omega'}=0)$ 
absorption rates~(\ref{eq:Wscatt}) can be expressed in terms of partial
cross sections
$W_{\lambda \to \lambda^{'}} \equiv W_{\lambda e^{-} \to \lambda^{'}
e^{-}} + W_{\lambda e^{+} \to \lambda^{'} e^{+}} \simeq  n_e
\sigma_{\lambda \to \lambda^{'}}$
%~\footnote{We considered the region $\omega \le 2m$.}
%
\begin{widetext}
\begin{eqnarray}
\sigma_{1 \to 1} = \frac{3}{16}\, \sigma_T \left
(\frac{B_e}{B} \right)^2 \frac{\omega^2}{m^2}\, \left [\frac{(2m +
\omega (1-u^2))(m + \omega + (m - \omega)u^2)}
{(\omega + m)^2 - \omega^2 u^2}\,  
%\nonumber
+  \frac{m}{\omega} (1-u^2) \ln \left (\frac{(\omega
+ m)^2 - \omega^2 u^2}{m^2} \right)\right ],
\label{eq:sigma11}
\end{eqnarray}
%
%\vspace*{-11mm}
%
\begin{eqnarray}
\nonumber \sigma_{2 \to 1} = \frac{3}{16}\, \sigma_T \left
(\frac{B_e}{B} \right)^2 \frac{q^2_{\mprp}}{m^2 \omega}\, Z_2 \,
\left [\frac{(2m \omega + q^2_{\mprl})(2m\omega^2 - (m - \omega
)q^2_{\mprl})(1-u^2)} {q^2_{\mprl} [(\omega + m)^2(1 - u^2) - q^2_{\mprp}
u^2]}\, 
%\nonumber
-  m \ln \left (\frac{(\omega + m)^2 (1-u^2) - q^2_{\mprp}
u^2}{(1-u^2) m^2} \right)\right ], 
\end{eqnarray}
%
%\vspace*{-18mm}
%
\begin{eqnarray} 
q^2_{\mprl} = \omega^2 -
q^2_{\mprp} u^2/(1 - u^2)\label{eq:sigma21}
\end{eqnarray}
%
%\vspace{-11mm}
%
\begin{eqnarray}
\sigma_{1 \to 2} = \frac{3}{4}\, \sigma_T  \left (  \frac{B_e}{B}
\right)^2 \frac{1}{\omega s_{1 \mprl}} \!\!\!\! \int
\limits_0^{\,\,\,\,\,\, (m-\sqrt{s_{1 \mprl}})^2\!/4m^2}
\!\!\!\!\!\!\! dz \left(1 + \frac{3}{2} \xi \,\frac{H(z)}{z} \right)
 \frac{(\omega+m)(s_{1 \mprl}-m^2)^2-4m^2z(m^2(\omega+m)+
(\omega+3m)s_{\mprl})}{\sqrt{(s_{1 \mprl}-m^2+4m^2z)^2 - 4 m^2
s_{1 \mprl}z}},
\nonumber 
\end{eqnarray}
%
%\vspace*{-1mm}
%
%
\begin{eqnarray}
s_{1 \mprl}& = &(\omega + m)^2 - \omega^2 u^2,
\label{eq:sigma12}
\end{eqnarray}
\begin{eqnarray}
%\!\!\!\!\!\!\!\!\!\!\!\!
\,\,\,\,\,&\sigma_{2 \to 2}& =  3\, \sigma_T \,
\frac{m}{\omega} \, Z_2 \, \int \limits_0^{\,\,\,\,\,\,
(\sqrt{s_{2\mprl}}-m)^2\!/4m^2}
\!\!\!\!\!\!\! \frac{dz}{\sqrt{(s_{2\mprl}-m^2+4m^2z)^2 - 16 m^2 s_{2\mprl}z}}
\times
%\nonumber
\label{eq:sigma22}
\\
&\times& \sum^2_{i=1}\frac{4m^4} {[(q \tilde \Lambda
q')^2 - \varkappa^2 (q \tilde \varphi q')^2]^2} \bigg [
4m^2 q^2_{\mprl}z (4m^2 - Q^2_{\mprl}) \left(1 - \frac{B_e}{B}\left
(\frac{q^2_{\mprp}}{4m^2}
+ z + \frac{3}{2} \xi \,H(z)\right ) \right)^2 +
\nonumber
\\ [3mm]
&+&\frac{1}{8m^4}\left (\frac{B_e}{B} \right)^2 \,\frac{q^2_{\mprp}(q \tilde \varphi q')^2}
{q^2_{\mprl} (-Q^2_{\mprl})}\,\left(1 + \frac{3}{2} \xi \,\frac{H(z)}{z} \right) \,
\left (4m^2 (q \tilde \Lambda q') + (q \tilde \Lambda
q')^2 - \varkappa^2 (q \tilde \varphi q')^2 \right )^2
\bigg ]_{\bigg |_{\,q'_z = q'_{zi}} \!\!\!\!\!\!\!\! \!\!\!\!\!\!\!\!
\omega' = \omega'_{i}}
\nonumber
\end{eqnarray}
%\\ [3mm]
\begin{eqnarray}
\omega'_{1,2}& = &\frac{1}{2s_{2\mprl}} \left \{(\omega + m)
(s_{2\mprl}-m^2 + 4m^2 z) \pm  \sqrt{q^2_{\mprp}} \frac{u}{\sqrt{1-u^2}}\,
\sqrt{(s_{2\mprl}-m^2+4m^2z)^2 - 4 m^2 s_{2\mprl}z}\right \},
\nonumber
\\
q'_{z1,2}& = &\frac{1}{2s_{2\mprl}} \left \{\sqrt{q^2_{\mprp}} \frac{u}{\sqrt{1-u^2}}\,
(s_{2\mprl}-m^2 + 4m^2 z) \pm (\omega + m)
\sqrt{(s_{2\mprl}-m^2+4m^2z)^2 - 4 m^2 s_{2\mprl}z}\right \},
\nonumber
\\
s_{2\mprl}& = &(\omega + m)^2 - q^2_{\mprp} \frac{u^2}{1-u^2},
\nonumber
\end{eqnarray}
\end{widetext}

\begin{figure}
\includegraphics[width=8cm]{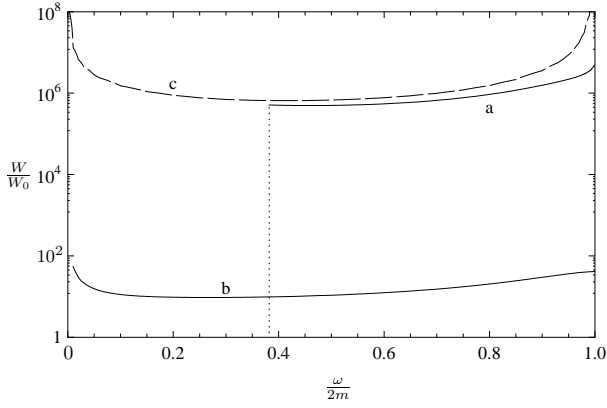}\vspace*{-2mm} \caption{
The dependence of the total photon absorption rate $W_{2 \to 2} + W_{2 \to 1}$
%of the channels $\gamma_2 e \to \gamma_2 e$  and  $\gamma_2 e \to \gamma_1 e$  
on energy of the initial photon in
strong magnetic field $B/B_e \, = \, 200$ at $T \,=\,1$ MeV -- {\it a} and
$T \,=\,50$ keV -- {\it b}.
The long dashed line  {\it c} corresponds to the Compton scattering
absorption rates without taking into account photon dispersion and
wave function renormalization at $T \, = \, 1$ MeV. Here, $\theta = \pi/2$.
 \label{fig:com12_21}}
\end{figure}

\noindent where $\sigma_T = \frac{8\pi}{3} \frac{\alpha^2} {m^2}$ is the Thompson cross section, $\xi$ is the parameter defined in Sec. IV, $q^2_{\mprp}$ is the root of equation $q^2_{\mprp}= (1-u^2)\, [\omega^2 -
\P^{(2)}(\omega^2 - q^2_{\mprp} u^2/(1-u^2))]$ and the number of
electron (positron) density in a strongly magnetized and
charge-symmetric rarefied plasma is defined by~(\ref{eq:ne}).
To verify the result obtained we have calculated the corresponding cross sections in low-energy limit($\omega \ll m$):
\begin{eqnarray}
\sigma_{1 \to 1} &\simeq& \frac{3}{4}\, \sigma_T  \left ( \frac{B_e}{B}
\right )^2 \frac{\omega^2}{m^2}, \label{eq:sigma11sT}
\end{eqnarray}
%\\
\begin{eqnarray}
\sigma_{1 \to 2} &\simeq& \frac{1}{4}\, \sigma_T  \left ( \frac{B_e}{B}
\right )^2 \frac{\omega^2}{m^2}\, (1 + \xi), \label{eq:sigma12sT}
\end{eqnarray}
%\\
\begin{eqnarray}
\sigma_{2 \to 1} &\simeq& \frac{3}{4}\, \sigma_T \left ( \frac{B_e}{B}
\right )^2 \frac{(\omega - \omega_{pl})^2}{m^2} \, (1 + \xi)  u^2 \, \Theta(\omega - \omega_{pl}) ,\nonumber \\
\label{eq:sigma21sT}
\end{eqnarray}
\begin{eqnarray}
\sigma_{2 \to 2} &\simeq&  \frac{\sigma_T}{1+\xi} \bigg \{(1-u^2)
%\right.
\nonumber
%\left. 
\left [1- \frac{1}{2}\, \left ( \frac{B_e}{B} \right )
\left ( \frac{\omega}{m} \right )^2 \left ( \frac{9}{5} - u^2 \right ) \right ]
%\right.
\nonumber
\\[2mm]
&+& %\left. 
\left ( \frac{B_e}{B} \right )^2 \frac{(\omega -
\omega_{pl})^2}{4 m^2} u^2 \bigg \} \Theta(\omega - \omega_{pl}),
\label{eq:sigma2sT}
\end{eqnarray}

\begin{figure}
\includegraphics[width=8cm]{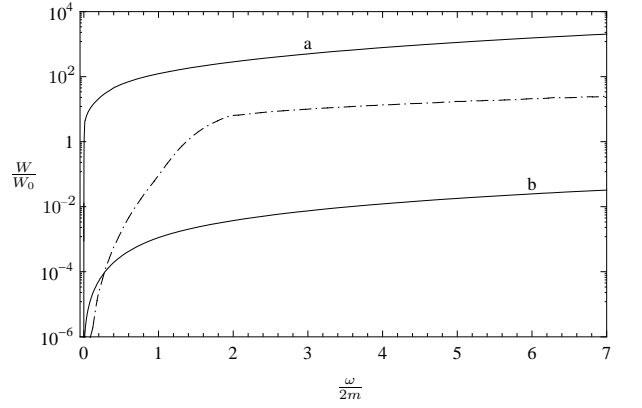}\vspace*{-2mm} \caption{
The dependence of the total photon absorption rate $W_{1\to 1} + W_{1\to 2}$
on energy of the initial photon in
strong magnetic field $B/B_e \, = \, 200$ at $T \,=\,1$ MeV -- {\it a} and
$T \,=\,50$ keV -- {\it b}. The dashed-dotted line corresponds to the total probability of
photon splitting $W_{1 \to 12} +W_{1 \to 22}$ 
at $T \, = \, 50$ keV. Here, $\theta = \pi/2$.
 \label{fig:com11_22}}
\end{figure}

One can see that the presence of magnetized plasma slightly
influences on the process cross sections in this limit. Moreover,
the corrections connected with photon dispersion and wave function
renormalization are significant only for  $\xi \sim 1$, i.e., when
the magnetic field is rather strong $B \sim 10^3 B_e$. In the case $\xi
\ll 1$ which is relevant for the models of magnetar magnetosphere
emission the formulas (\ref{eq:sigma11sT} -- \ref{eq:sigma21sT})
coincide with the well-known result~\cite{Herold:1979}. However, the cross 
section~(\ref{eq:sigma2sT}) contains the extra term $\sim B_e/B$ in comparison with 
result~\cite{Herold:1979}. This term 
arises from the series  expansion of exponents in the amplitude~(\ref{eq:Mcompt}) 
(see also~(\ref{eq:M22}))  in terms of  magnetic field strength.

For the numerical analysis of photon absorption rates under hot 
plasma conditions ($T \sim m$) it is convenient to make 
integration in~(\ref{eq:Wscatt}) over $p_z$. Then for  
$W_{\lambda  \to \lambda'}$ we obtain the following simplified expression:
%
%\begin{widetext}
\begin{eqnarray}
\nonumber
&&W_{\lambda  \to \lambda'} = \frac{eB Z_{\lambda}}
{4 (2\pi)^4 \omega_{\lambda}}
\int \frac{d^3k'}{\varkappa |Q_{\mprl}^2| \omega_{\lambda'}}
\mid {\cal M_{\lambda \to \lambda^{'} }}\mid^2 Z_{\lambda'}
\\
[0.1in]
\nonumber
&&\times (1+f_{\omega'}) \bigg \{{\cal F}_5 \left [-\frac{1}{2T} (Q_0 - Q_z \varkappa),
\frac{1}{2T} (Q_0 + Q_z \varkappa) \right ]
\\
[0.1in]
\label{eq:Wscatnum}
&& \times \Theta(Q_z) +
 (Q_z \to -Q_z) \bigg \}\, ,
\end{eqnarray}
%\end{widetext}
\noindent where ${\cal F}_5 (x,y) = [1+\exp{(x)}]^{-1} [1+\exp{(-y)}]^{-1}$.

\begin{figure}
\includegraphics[width=8cm]{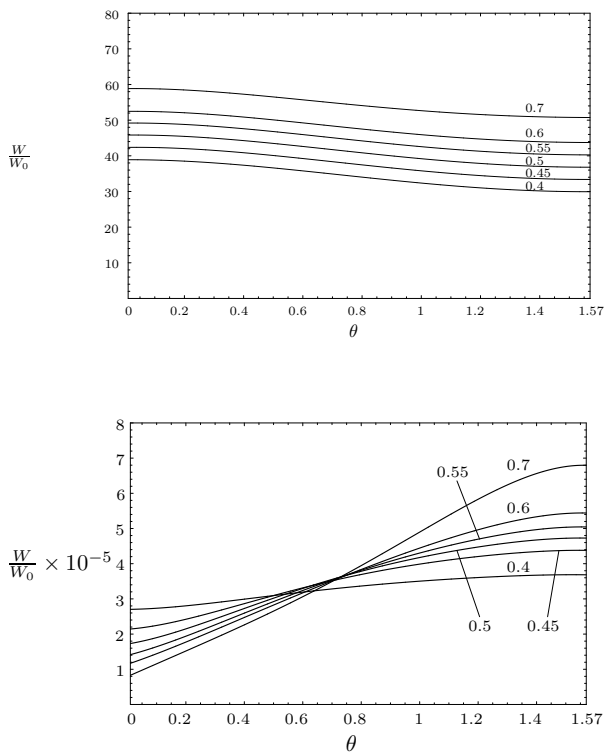}\vspace*{-2mm} \caption{
The
dependence of the absorption rates  $W_{1 \to 1}$ (upper panel) and  $W_{2 \to 2}$
 (lower panel) on the angle
between initial photon momentum and magnetic field direction at
different initial photon energies ($B/B_e \, = \, 200, T \, = \, 1
\mbox{MeV}$). The numbers above the curves correspond to the values of
the ratio $\omega / 2 m$. \label{fig:compt11_22}}
\end{figure}

\begin{figure}
\includegraphics[width=7.1cm]{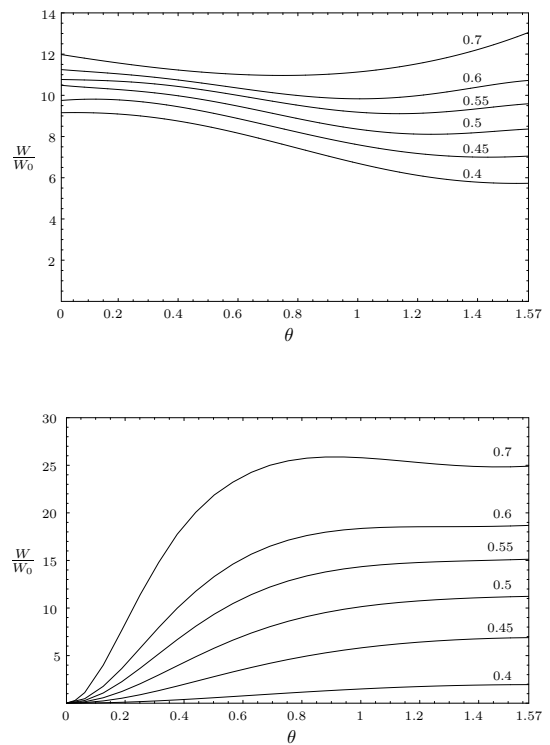}\vspace*{-2mm} \caption{
The
dependence of the absorption rates of the channels $W_{1 \to 2}$
  (upper panel) and  $W_{2 \to 1}$ (lower panel) on the angle
between initial photon momentum and magnetic field direction at
different initial photon energies ($B/B_e \, = \, 200, T \, = \, 1
\mbox{MeV}$). The numbers above the curves correspond to the values of
the ratio $\omega / 2 m$. \label{fig:compt12_21}}
\end{figure}

The results of  numerical calculations  
are presented  in Figs.
\ref{fig:com12_21}-\ref{fig:compt12_21}. In the
 Figs. \ref{fig:com12_21} and \ref{fig:com11_22} one can see that photon
absorption rates corresponding to Compton scattering are the fast
increasing functions of temperature. At the same time, the channels
with initial photon of mode 1 and mode 2 have different character of
the absorption coefficient  energy dependence. 

As shown in the
Fig.\ref{fig:com12_21}                                               
the total absorption rate for the reactions $\gamma_2
e^{\pm} \to \gamma_2 e^{\pm}$  and $\gamma_2 e^{\pm} \to \gamma_1
e^{\pm}$ have threshold $(\omega = \omega_{pl})$. It is  caused by mode 2 dispersion
relation  and indicates the fact that the
electromagnetic wave corresponding to mode 2 cannot propagate with
energy below $\omega_{pl}$.

On the other hand, the electromagnetic wave, corresponding to the 2-mode 
quickly attenuates in the region $\omega \geqslant 2m$ due to the 
process $\gamma_2 \to e^+e^-$.    
We would like to note that in the vicinity of the pair
creation threshold taking into account the wave function renormalization
and photon dispersion becomes very important and defines the
processes rates' dependencies on energy, temperature and magnetic
field. It is well seen from the
comparison of the solid  and long dashed lines {\it a} and {\it c} (Fig. \ref{fig:com12_21}) 
calculated with and without taking into account photon dispersion
and wave function renormalization, correspondingly. 

The energy dependence of
total rates $W_{1 \to 1} + W_{1 \to 2}$
is depicted in the
Fig. \ref{fig:com11_22}. One can see
the fast increase of absorption coefficients at low energies and
rather slow dependence at $\omega \gtrsim 2 m$.  Such behavior 
 indicates the possibility of the 1-mode photons
efficient diffusion in the emission region whereas 2-mode  seemed
to be trapped.

It is interesting also to consider the angle distributions of
the photon absorption rates wasn't analysed before in~\cite{RCh09}
 (see Figs.~\ref{fig:compt11_22} and~\ref{fig:compt12_21}).
It is seen, that in the hot plasma the angle dependence of 
$W_{1\to 1}$  and $W_{1\to 2}$ is close to isotropic distribution. Recall,  
that in the low-temperature limit, the same angle distribution is strictly 
isotropic (see~(\ref{eq:sigma11sT}) and~(\ref{eq:sigma12sT})). 
In contrast, the absorption rates $W_{2\to 1}$ and $W_{2\to 2}$ strongly 
depend on angle. They are minimized at $\theta = 0$ and have maximum when 
2-mode initial photon propagates across magnetic field direction.

\section{Discussions}
\label{Sec:6}

In the models of soft gamma repeaters spectrum formation the
dependence of photon absorption rates on energy  and temperature
plays an important role. It could influence  the shape of emergent
spectrum and define the temperature profile in the emission region
during bursts in SGRs~\cite{Thompson:1995, Lyubarsky:2002,Thompson:2001}.

The previous investigations of the radiation transfer problem in strongly
magnetized  plasma have shown that along with the Compton scattering
process the photon splitting $\gamma \to \gamma \gamma$ could play a
significant role as a mechanism of photon
production~\cite{Thompson:1995, Thompson:2001}.
In Sec. IV it was shown that at the
temperature $T \sim m$  
in kinematical region $\omega \le 2 m $ the main photon splitting
process is $\gamma_2 \to \gamma_1 \gamma_1$ forbidden in pure
magnetic field.  
However, the comparison of Fig.~\ref{fig:W211} and Fig.~\ref{fig:com12_21} 
shows that the rate of this 
process is much slower than the Compton scattering one $W_{2 \to 1} + W_{2 \to 2}$.
Nevertheless,  it
could be an effective photon production mechanism at temperatures
under consideration. 

The total probability of the channels $\gamma_1 \to
\gamma_1 \gamma_2$ and $\gamma_1 \to \gamma_2 \gamma_2$ increases
with temperature falling  and becomes comparable and even larger than
the total Compton scattering rate $W_{1 \to 1} + W_{1 \to 2}$.
 As shown in
Fig.~\ref{fig:com11_22} the process of photon splitting (dashed-dotted line) strongly
dominates over Compton scattering at $T=50$ keV. 

It was claimed
previously that the effect of strongly magnetized cold plasma on
photon splitting is not pronounced and the vacuum approximation can
be used in the most calculations~\cite{Bulik:1997,Elmfors:1998}. 
 Our analysis shows  that in the presence of hot plasma the process of photon
splitting could be not only an intensive source of photon production
but also an effective absorption mechanism.

\begin{figure}
\centerline{\includegraphics[width=7.9cm]{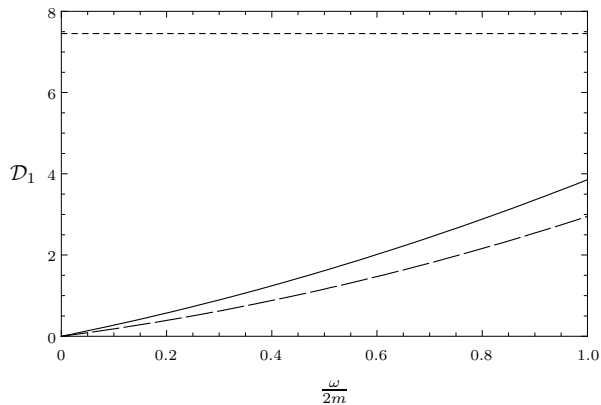}}
\vspace*{-2mm} \caption{
The diffusion coefficient ${\cal D}_1 = \omega^2 W_0 D_1/(2m)^2$, where $D_1$ is
defined by (\ref{eq:diff1}), as a function of the ratio $\omega/2m$  at
$T = 1$ MeV and $B = 200 B_e$. The solid and long dashed lines correspond
to diffusion coefficient
with and without taking into account photon dispersion and
wave function renormalization. The approximation 
(\ref{eq:diff2}) is depicted by  the short dashed line.}
\label{fig:diff}
\end{figure}
%

%\newpage
Let 
 us illustrate this  fact in the framework of the magnetar model
of SGR burst. It  is known that the radiation transfer in the
magnetically trapped plasma may be described  as ô diffusion of the
1-mode photons whereas 2-mode photons are
locked~\cite{Thompson:1995, Lyubarsky:2002,Thompson:2001}.
The last circumstance is
connected 
 especially with weak dependence of the 2-mode absorption
coefficient on photon energy (see Fig.~\ref{fig:com11_22}).
We can approximatively describe the radiation transfer 
under the considered
conditions via a diffusion equation 
(see  for example~\cite{Nagel})\footnote{We consider plane-parallel geometry when
the temperature gradient and magnetic field are directed along the z-axis}:
\begin{eqnarray}
\frac{\partial n^\omega_\lambda}{\partial t} -
\frac{\partial}{\partial z} \left (D_\lambda \frac{\partial
n^\omega_\lambda}{\partial z}\right ) =
Q^\omega_\lambda \, ,
\label{eq:diffeq1}
\end{eqnarray}
\noindent
where $n^\omega_\lambda$ is the photon number of density for $\lambda =1,2$ modes,
$Q^{\omega}_\lambda$ is the source of $\lambda$-mode photon,
% creation and absorption rates,
%
\begin{eqnarray}
 D_\lambda (\omega, z)=  \int \frac{d\Omega}{4\pi}
 \ell_\lambda (\theta, \omega, z)
\cos^2{\theta}\, 
\label{eq:diff1}
\end{eqnarray}
\noindent is the diffusion coefficient and $D_1 \gg D_2$,
\begin{equation}
\ell_\lambda = \left[\,\sum \limits_{\lambda'=1}^2
W_{\lambda \to \lambda'}  + \sum \limits_{\lambda',\lambda''=1}^2
\left( W_{\lambda \to \lambda^{'} \lambda^{''}} + W_{\lambda
\lambda^{'} \to \lambda^{''}} \right) \right]^{-1} \, 
\label{eq:path}
\end{equation}
\noindent is the mode $\lambda$ photon free path.

Note that in the magnetar model of SGR burst~\cite{Thompson:1995, Thompson:2001} in
the analysis of radiation transfer Herold's approximation of Compton scattering cross 
sections~(\ref{eq:sigma11sT}) and~(\ref{eq:sigma12sT})  
was used. In this case the 1-mode photon free path can be written as
\begin{equation}
\ell_1^H = 3 D_1 \simeq \frac{1}{n_e \sigma_T} \,
\left (\frac{m}{\omega} \right )^2 \left (\frac{B}{B_e} \right )^2 \, .
\label{eq:diff2}
\end{equation}

In the Fig.~\ref{fig:diff}  we demonstrate the importance of taking into account 
dispersion and wave function renormalization  of photons in the radiation transfer 
problem. From the analysis above it follows that in the hot plasma $T\sim m$ the 
Compton scattering gives the main contribution to the photon free path and, as a 
consequence, to the diffusion coefficient. It is seen that photon 
dispersion and wave function renormalization become essential at initial photon 
energies, $\omega \gtrsim  m$ (compare solid and long dashed lines in the 
Fig.~\ref{fig:com11_22}). We can see also that approximation~(\ref{eq:diff2})
is not applicable in hot plasma (dashed line). 

\begin{figure}
\centerline{\includegraphics[width=8cm]{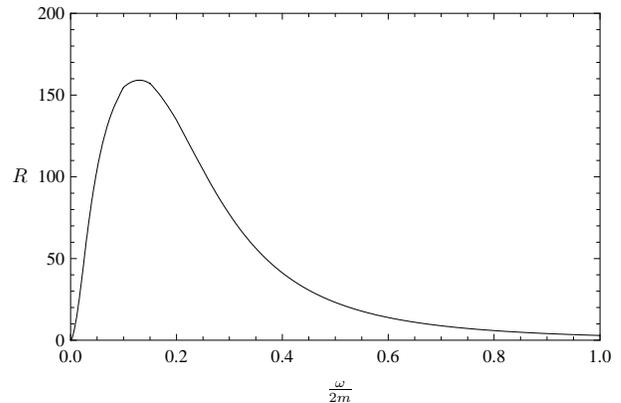}}
\vspace*{-2mm} \caption{The ratio $R=\ell_1/\ell_1^H$ as
a function of the photon energy at $B = 200 B_e$}
\label{fig:RC_CSp}
\end{figure}

To demonstrate that the photon splitting process can be considered not only as the source 
of the photons but also as an effective absorption mechanism we depict the ratio of 1-mode 
photon free path~(\ref{eq:path}) and approximation~(\ref{eq:diff2}), $R=\ell_1/\ell_1^H$ 
at $T=50$ keV and $\theta = \pi/2$ (Fig.~\ref{fig:RC_CSp}).  
 It is seen that taking into account the photon splitting contribution leads to the 
essential increase of 1-mode photon free path in comparison to the  commonly used 
approximation~(\ref{eq:diff2}) in a wide  range of photon energies.

 The solution
of the diffusion Eq. (\ref{eq:diffeq1})  is out of the scope of
this article. However, we would like to note that the more detailed
analysis of the radiation transfer needs the consistent solution of
the Boltzmann equation for the photon occupation number and radiative
transfer equation in the wide range of temperatures (10 keV
$\lesssim T \lesssim $ 1 MeV).

%\vspace{-3mm}
\section{Conclusions}
\label{Sec:7}

%\vspace{-3mm}
In conclusion, we have investigated the influence of the strongly
magnetized hot  plasma on the photon splitting (merging) and Compton scattering 
processes  taking
into account the photon dispersion and large radiative corrections.
The partial amplitudes and polarization selection rules for the process of photon splitting 
 were obtained.
The obtaining results show that  plasma influence modifies 
the polarization selection rules in comparison to the pure magnetic field. 
In particular, the new splitting channel $\gamma_2 \to \gamma_1 \gamma_1$,    
forbidden without plasma, is allowed.
On the other hand, 
the presence of plasma 
suppresses the probabilities of channels $\gamma_1 \to \gamma_1 \gamma_2$ 
and $\gamma_1 \to \gamma_2 \gamma_2$ in comparison to the pure magnetic field. 

In addition,  it was
found that in hot plasma radiation transfer mainly occurs by means
of 1-mode  photon diffusion whereas the 2-mode is trapped. 
 The comparison
of the photon splitting (merging) processes and Compton scattering shows that
the influences of these reactions on the 1-mode radiation transfer
are competitive in rarefied plasma $(T \ll m)$.
As a result, it could lead to the 
modification in the mechanism of 
the spectra formation of SGR and AXP.

\vspace{-2mm}
\bigskip

{\bf Acknowledgements}

We are grateful to N.V. Mikheev and A.V. Kuznetsov 
for stimulating discussions and valuable comments.

The study was performed within the State Assignment for Yaroslavl
University Project No.~2.4176.2011, and was supported in part by the
Russian Foundation for Basic Research Project No.~11-02-00394-a.

%%%%%%%%%%%%%%%%%%%%%%%%%%%%%%%%%%%%%%%%%%%%%%%%%%%%%%%%%%%%%%%%%%%%%%%%%
%\newpage

%%%%%%%%%%%%%%%%%%%%%%%%%%%%%%%%%%%%%%%%%%%%%%%%%%%%%%%%%%%%%%%%%%%%%%%%%
\end{document}